\RequirePackage{docswitch}
\def\flag{mnras}
\setjournal{\flag}
\documentclass[\docopts]{\docclass} 

\usepackage{lsstdesc_macros}
\usepackage{graphicx}
\usepackage{hyperref}
\usepackage{url}
\usepackage{natbib}
\usepackage[frozencache]{minted}
\usepackage{tikz}
\usepackage{amssymb}
\usepackage[normalem]{ulem}
\usepackage{xcolor}
\graphicspath{{./}{./figures/}}

\usetikzlibrary{arrows.meta}
\usetikzlibrary{shapes.geometric}
\usetikzlibrary{shapes.misc}
\usetikzlibrary{positioning}
\definecolor{TheoryColor}{HTML}{38761d}
\definecolor{DataOpsColor}{HTML}{1155cc}
\definecolor{GCColor}{HTML}{85200c}
\definecolor{CosmoColor}{HTML}{351c75}
\definecolor{UtilsColor}{HTML}{0c433d}
\definecolor{GCDataColor}{HTML}{7f6000}
\definecolor{SupportColor}{HTML}{000000}

\definecolor{TheoryColorfill}{HTML}{63d297}
\definecolor{DataOpsColorfill}{HTML}{c9daf8}
\definecolor{GCColorfill}{HTML}{f4cccc}
\definecolor{CosmoColorfill}{HTML}{d9d2e9}
\definecolor{UtilsColorfill}{HTML}{d9ead3}
\definecolor{GCDataColorfill}{HTML}{fff2cc}
\definecolor{SupportColorfill}{HTML}{cccccc}

\definecolor{faintred}{HTML}{ea9999}
\definecolor{PipelineColor}{HTML}{85200c}
\definecolor{PipelineColorfill}{HTML}{f4cccc}
\definecolor{DataProductColor}{HTML}{38761d}
\definecolor{DataProductColorfill}{HTML}{63d297}
\tikzset{%
  >={Latex[width=2.5mm,length=2mm]},
    package/.style = {rectangle, draw=black,
        minimum width=5cm, minimum height=2.7cm,
        text width=4.8cm,
        text centered,
        line width=1.5pt,
        font=\sffamily},
    module/.style = {package, rounded corners},
    pipe/.style = {
        draw=PipelineColor,
        fill=PipelineColorfill,
        minimum width=4cm, minimum height=1cm,
        text width=3.8cm,
        text centered,
        line width=1.5pt,
        font=\sffamily},
    pipeline/.style = {rectangle, pipe},
    dataproduct/.style = {
        rounded corners,
        pipe,
        minimum width=.5cm,
        text width=3.cm,
        draw=DataProductColor,
        fill=DataProductColorfill,
        },
}
\newcommand{\diagText}[1]{\texttt{#1}}
\newcommand{\diagTitle}[3][black]{
    \scalebox{1.4}{
        \textbf{
            \color{#1}
            \hypersetup{linkcolor=.}
                \hyperref[#3]{#2}
        }
    }
}

\def\DataOps{{\color{DataOpsColor}\textbf{DataOps}}}
\def\GCData{{\color{GCDataColor}\textbf{GCData}}}
\newcommand{\diagram}{
\begin{tikzpicture}[node distance=3cm,
    every node/.style={fill=white, font=\sffamily}, align=center]
   \node(DataOps)[package, draw=DataOpsColor, fill=DataOpsColorfill]{
        \diagTitle[DataOpsColor]{DataOps (\S\ref{sec:dataops})}{sec:dataops_ops}\\
        \diagText{Compute desired quantities from measurements (shear components, profiles)}
    };
   \node(Theory)[package, 
   below=.5cm of DataOps,
   draw=TheoryColor, fill=TheoryColorfill]{
        \diagTitle[TheoryColor]{Theory (\S\ref{sec:theory})}{sec:theory}\\
        (cluster\_toolkit, CCL, NumCosmo)\\
        \diagText{Theoretical prediction for lensing properties (surface density, shear, magnification)}
    };
   \node(GC)[module,
   right=1cm of Theory,
   draw=GCColor, fill=GCColorfill]{
        \diagTitle[GCColor]{GalaxyCluster (\S\ref{sec:galaxycluster})}{sec:galaxycluster}\\
        \diagText{Cluster information (ID, ra, dec, z), \DataOps\ functionality, galaxy source catalog (\GCData), shear profiles (\GCData)}
    };
   \node(Utils)[module,
   left=1cm of DataOps,
   draw=UtilsColor, fill=UtilsColorfill]{
        \diagTitle[UtilsColor]{Utils (\S\ref{sec:utils})}{sec:utils}\\
        \diagText{Auxiliary functions (binning, conversions, simple computations)}
    };
   \node(Cosmo)[package,
   left=1cm of Theory,
   draw=CosmoColor, fill=CosmoColorfill]{
        \diagTitle[CosmoColor]{Cosmology (\S\ref{sec:cosmology})}{sec:cosmology}\\
        (astropy, CCL, NumCosmo)\\
        \diagText{Cosmological operations\ (distances, parameter evolution)}
    };
   \node(GCData)[module,
   right=1cm of DataOps,
   draw=GCDataColor, fill=GCDataColorfill]{
        \diagTitle[GCDataColor]{GCData (\S\ref{sec:gcdata})}{sec:gcdata}\\
        \diagText{Astropy table with added functionality to check cosmology signature}
    };
   \node(Support)[package, minimum height=1.7cm, minimum width=17cm, text width=16.8cm,
   below=.5cm of Theory,
   draw=SupportColor, fill=SupportColorfill]{
        \diagTitle[SupportColor]{Support (\S\ref{sec:support})}{sec:support}\\
        \diagText{
        \hypersetup{linkcolor=.}
        Operations to test and facilitate the use of clmm (\hyperref[sec:mock_data]{mock data},  \hyperref[sec:samplers]{parameters fitting/sampling})}
    };
 \draw[line width=2pt, color=red, ->](Theory) -- (Support);
 \draw[line width=2pt, color=red, ->](Utils.west) .. controls +(down:10mm) and +(left:1cm) .. (Support.west);
 \draw[line width=2pt, color=red, ->](GCData.east) .. controls +(down:10mm) and +(right:1cm) .. (Support.east);
  \draw[line width=2pt, color=red, ->](GCData) -- (GC);
  \draw[line width=2pt, color=red, ->](Utils) -- (DataOps);
  \draw[line width=2pt, color=red, ->](DataOps) -- (GC);
  \draw[line width=2pt, color=red, ->](Theory) -- (GC);
  \draw[line width=2pt, color=red, ->](GCData) -- (DataOps);
  \draw[line width=2pt, dashed, color=faintred, ->](Cosmo) -- (Theory);
  \draw[line width=2pt, dashed, color=faintred, ->](Cosmo) -- (DataOps);
  \end{tikzpicture}
}
\newcommand{\diagramPipeline}{
    \begin{tikzpicture}[node distance=3cm,
        every node/.style={fill=white, font=\sffamily}, align=center]
    \node(ExtData)[dataproduct]{External Data (Elipticity measurements, ...)};
   \node(Data)[pipeline, below=.5cm of ExtData]{CLMM (Data modules*)};
   \node(Theory)[pipeline, right=.5cm of Data]{CLMM (Theory modules*)};
    \node(DataShear)[dataproduct, below=0.2 of Data]{Shear profile measured ($\Delta\Sigma$, $g_t$, ...)};
    \node(TheoShear)[dataproduct, below=0.2 of Theory]{Theoretical prediction of shear profile ($\Delta\Sigma$, $g_t$, ...)};
    \node(Like)[pipeline, below right=0.5 and -1.4cm of DataShear]{Likelihood+Parameter constraining (CLMM optional**)};
    \node(Par)[dataproduct, below=0.2 of Like]{Parameter constraints (Mass, concentration, ...)};
  \draw[line width=2pt, dashed, color=faintred, ->](ExtData) -- (Data);
  \draw[line width=1pt, color=red, -](Data) -- (DataShear);
  \draw[line width=1pt, color=red, -](Theory) -- (TheoShear);
  \draw[line width=2pt, dashed, color=faintred,  ->](DataShear) -- (Like);
  \draw[line width=2pt, dashed, color=faintred,  ->](TheoShear) -- (Like);
  \draw[line width=1pt, color=red, -](Like) -- (Par);
  \end{tikzpicture}
}

\newcommand{\clmm}{\code{CLMM}}
\newcommand{\ct}{\code{CT}}
\newcommand{\ccl}{\code{CCL}}
\newcommand{\nc}{\code{NC}}
\newcommand{\astropy}{\code{Astropy}}

\newcommand{\collab}[1]{\textsc{#1}~}
\newcommand{\lsst}{\collab{LSST}}
\newcommand{\desc}{\collab{LSST-DESC}}

\newcommand{\be}{back-end}

\newcommand\redout{\bgroup\markoverwith
{\textcolor{red}{\rule[.5ex]{2pt}{0.4pt}}}\ULon}

\begin{document}
\title{
\clmm: a LSST-DESC Cluster weak Lensing Mass Modeling library for cosmology
}

\maketitlepre
\author[LSST Dark Energy Science Collaboration]{
\parbox{\textwidth}{
\Large
M.~Aguena,$^{1,2}$
C.~Avestruz,$^{3,4}$\thanks{cavestru@umich.edu},
C.~Combet,$^{5}$\thanks{celine.combet@lpsc.in2p3.fr}
S.~Fu,$^{6}$
R.~Herbonnet,$^{7}$
A.I.~Malz,$^{8}$\thanks{aimalz@astro.ruhr-uni-bochum.de}
M.~Penna-Lima,$^{9,1}$
M.~Ricci,$^{2}$
S.~D.~P.~Vitenti,$^{10,1}$
L.~Baumont,$^{7}$
H.~Fan,$^{7}$
M.~Fong,$^{11}$
M.~Ho,$^{12}$
M.~Kirby,$^{13}$
C.~Payerne,$^{5}$
D.~Boutigny,$^{2}$
B.~Lee,$^{14}$
B.~Liu,$^{6}$
T.~McClintock,$^{15}$
H.~Miyatake,$^{16,17,18}$
C.~Sif\'on,$^{19}$
A.~von der Linden,$^{7}$
H.~Wu,$^{20}$
and M.~Yoon$^{8,21}$
\begin{center} (The LSST Dark Energy Science Collaboration) \end{center}
}
\vspace{0.4cm}
\\
\parbox{\textwidth}{
$^{1}$ Laboratorio Interinstitucional de e-Astronomia, Rua General Jos\'e Cristino 77, Rio de Janeiro, RJ, 20921-400, Brasil\\
$^{2}$ Univ. Grenoble Alpes, Univ. Savoie Mont Blanc, CNRS, LAPP, 74000 Annecy, France\\
$^{3}$ Department of Physics, University of Michigan, 450 Church St, Ann Arbor, MI 48109-1040, U.S.A.\\
$^{4}$ Leinweber Center for Theoretical Physics, University of Michigan, 450 Church St, Ann Arbor, MI 48109-1040\\
$^{5}$ Univ. Grenoble Alpes, CNRS, LPSC-IN2P3, 38000 Grenoble, France\\
$^{6}$ Department of Physics, Brown University, 182 Hope Street, Box 1843, Providence, RI 02912, USA\\
$^{7}$ Department of Physics and Astronomy, Stony Brook University, Stony Brook, NY 11794, USA\\
$^{8}$ Ruhr-University Bochum, Astronomical Institute, German Centre for Cosmological Lensing, Universitaetsstr. 150, 44801 Bochum, Germany\\
$^{9}$ Universidade de Brasília, Instituto de Física, Caixa Postal 04455, Brasília, DF, 70919-970, Brazil\\
$^{10}$ Departamento  de  Física,  Universidade  Estadual  de  Londrina, Rod.   Celso  Garcia  Cid,  Km  380,  86057-970,  Londrina,  Paran\`a,  Brazil\\
$^{11}$ Department of Physics, The University of Texas at Dallas, Richardson, Texas 75080, USA \\
$^{12}$ McWilliams Center for Cosmology, Department of Physics, Carnegie Mellon University, Pittsburgh, PA 15213, USA\\
$^{13}$ Department of Physics, University of Arizona, Tuscon, AZ 85721, USA\\
$^{14}$ School for the Talented and Gifted at Townview, Dallas, Texas, USA\\
$^{15}$ Brookhaven National Laboratory, Upton, NY 11973, USA\\
$^{16}$ Kobayashi-Maskawa Institute for the Origin of Particles and the Universe (KMI), Nagoya
University, Nagoya, 464-8602, Japan\\
$^{17}$ Jet Propulsion Laboratory, California Institute of Technology, Pasadena, CA 91109, USA\\
$^{18}$ Kavli IPMU (WPI), UTIAS, The University of Tokyo, Chiba 277- 8583, Japan\\
$^{19}$ Instituto de F\'isica, Pontificia Universidad Cat\'olica de Valpara\'iso, Casilla 4059, Valpara\'iso, Chile\\
$^{20}$ Department of Physics, Boise State University, Boise, ID 83725, USA\\
$^{21}$ Department of Astronomy, Yonsei University, Yonsei-ro 50, Seoul, Republic of Korea\\
 }
}

\begin{abstract}
  We present the v1.0 release of \clmm, an open source \textit{Python} library for the estimation of the weak lensing masses of clusters of galaxies.  
  \clmm\ is designed as a standalone toolkit of
  building blocks to enable end-to-end analysis pipeline validation for upcoming cluster cosmology analyses such as the ones that will be performed by the LSST-DESC. 
  Its purpose is to serve as a flexible, easy-to-install and easy-to-use interface for both weak lensing simulators and observers and can be applied to real and mock data to study the systematics affecting weak lensing mass reconstruction. 
  At the core of \clmm\ are routines to model the weak lensing shear signal given the underlying mass distribution of galaxy clusters and a set of data operations to prepare the corresponding data vectors. The theoretical predictions rely on existing software, used as backends in the code, that have been thoroughly tested and cross-checked.
  Combined, theoretical predictions and data can be used to constrain the mass distribution of galaxy clusters as demonstrated in a suite of example Jupyter Notebooks shipped with the software and also available in the extensive online documentation.
\end{abstract}

\dockeys{galaxies: clusters: general; gravitational lensing: weak; software: public release}

\maketitlepost


\section{Introduction}
\label{sec:intro}

The Vera C. Rubin Observatory is an astronomical observatory, hosting a 8.4-m optical telescope and currently under construction at Cerro Pach\'on in Chile. 
During the first ten years of its operations, the Rubin Observatory will perform the Legacy Survey of Space and Time (\lsst).  
The survey will provide some of the deepest high-quality optical imaging to date to determine the colours and morphological properties of very distant galaxies over half of the sky \citep{lsst2012}. 

Data from \lsst\ enables five main dark energy probes, that are central to the work of the Dark Energy Science Collaboration (\desc): weak gravitational lensing, large scale structure, type Ia supernovae, strong gravitational lensing, and galaxy clusters \citep{2018arXiv180901669T}.
In particular, \lsst\ will probe the spatial density, distribution, and masses of galaxy clusters as a function of redshift. 
These galaxy cluster measurements will provide one of the strongest constraints on the growth of structure \citep{dodelson16}. 
Cosmological simulations of dark matter have shown that structures grow hierarchically.  Smaller distributions of matter, known as dark matter {\it halos}, merge to form larger {\it halos}. 
The number density of structures at a given mass, the halo mass function, therefore decreases with increasing mass. 
Galaxy clusters are the very end product of merging systems, and relatively rare objects. 
Their rarity and mass is the key to their role as powerful cosmological probes. 

Observational cluster cosmology compares the observed galaxy cluster number density to theoretical halo mass function predictions to determine cosmological parameters.  
To do so, observers need to find galaxy clusters and determine cluster masses. Observers use baryonic observables to detect galaxy clusters, such as through cluster member galaxies or via properties of the X-ray or SZ signal in clusters.  
However, these observables either do not directly provide estimates of total cluster mass, which is dominated by the dark matter halo, or rely on assumptions on the dynamical state of the cluster, resulting in a biased relation between the underlying total mass of a galaxy cluster and galaxy cluster observables (e.g., \citealp{hensonetal17,ansarifardetal20}). 

Cluster cosmology requires high precision and accuracy of the scaling relation between cluster observable and mass. 
Given the steepness of the halo mass function at the cluster scales, uncertainties in the cluster mass estimates lead to large uncertainties in derived cosmological parameters. 
Cluster cosmology, in its current state, is {\it limited by systematics}.  
One approach has been to rely on simulations to calibrate for the mass-observable scaling relation \citep{piffarettiandvaldarnini08,nelsonetal14,ansarifardetal20}. 
Recent results by \citet{2020arXiv201001138T} have shown that the mass-observable relation can also be calibrated using additional wide-area optical imaging data. 
On the other hand, gravitational lensing directly probes the total mass of along a line of sight without the assumptions on the intrinsic dynamics of the cluster. 
Simulations show that mass estimates from the weak lensing signal are nearly unbiased \citep{beckerandkravtsov11,hensonetal17,herbonnetetal20}. 
Weak lensing therefore provides an anchor for the mass-observable relation.

Gravitational lensing mass estimates of galaxy clusters may be derived from shape measurements of background source galaxies.  
Matter along the line-of-sight, including matter in galaxy clusters, deflects the paths of light coming from source galaxies, thereby distorting the observed shapes of the light sources. 
{\it Weak} gravitational lensing refers to when the distortion of galaxy shapes (shear) is small.  
In these cases, the distortion is too small to be measured from individual source galaxies. 
Instead, we statistically determine the weak shear signal by averaging the shapes of a large number of background source galaxies  \citep[e.g.][]{bartelmannandschneider2001}.  
Therefore, the statistical precision of weak lensing is limited by the available number of source galaxies. 
The unprecedented depth of \lsst\ provides enormous potential for weak gravitational lensing mass measurements of galaxy clusters, but also puts strong requirements on their accuracy.

Weak lensing masses are not devoid of systematic effects and there are five primary contributors to systematic biases in cluster weak lensing analyses: selection effects from cluster finding algorithm (e.g., \citealp{dietrich14, sunayama20}, inaccurate shear measurements \cite[e.g.][]{metacal,mandelbaum18,kannawadi19,hernandez20}, contamination of the source galaxy sample with galaxies at or in front of the cluster \cite[e.g.][]{medezinski18,varga19}, imperfect knowledge of the redshift distribution of the source galaxies \cite[e.g.][]{applegate14,wright20}, and simplifying (incorrect) assumptions on the mass distribution of an individual or a sample of galaxy clusters \citep{corless07,meneghetti10,beckerandkravtsov11}. 
Accounting for all systematic effects is a very challenging task as recently seen from recent results of large optical surveys (e.g., \citealp{2020PhRvD.102b3509A} and \citealp{2020arXiv201212273L}).

To extract the most information from cluster count cosmology with \lsst, we therefore need to carefully quantify each of these  systematic effects 
and calibrate observed weak lensing mass estimates accordingly, incorporating cutting-edge methodologies as they are developed prior to and throughout the ten-year survey. 
Such an endeavour requires a self-consistent flexible analysis code usable for both observations and simulations. 

In this paper, we present \clmm\footnote{See the Github repository at \url{https://github.com/LSSTDESC/CLMM}.}, an open source Python library providing tools to compute weak lensing mass estimates, using mock data, simulations or observational data. 
The \clmm\ project has drawn contributions from different analysis working groups in \desc\ with a common goal to produce user-friendly code and pedagogical documentation\footnote{The documentation is available at \url{http://lsstdesc.org/CLMM/}, which includes installation instructions, a detailed user API, and a set of rendered jupyter notebooks that highlight the main functionalities of the code.}.  
Our software has been internally reviewed by the collaboration.

We organize this paper as follows. 
In Section \ref{sec:scope} we first describe the motivation and scope of \clmm. 
Next we show the structure of the code in Section \ref{sec:stucture}, and then we give examples of running the code in Section \ref{sec:examples}. 
We outline future directions in Section \ref{sec:discussion} and finally summarize this paper in Section \ref{sec:summary}. 
Additionally, in Appendix \ref{sec:api_appendix} we present a demonstration of \clmm's API.

\section{Development Approach}
\label{sec:scope}

\begin{figure}
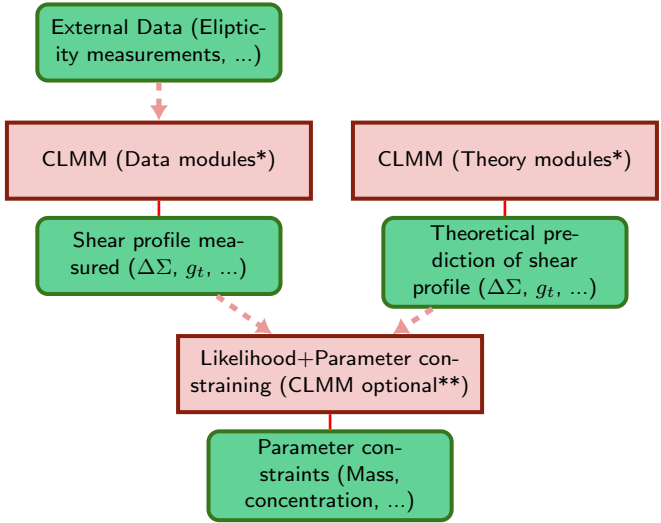

    \centering
    \diagramPipeline
    \caption{Pipeline diagram of \clmm's role in mass estimation. Operations are represented by red sharp edge rectangles and data products by green rounded corner rectangles. Straight lines represent direct products (outputs) and dashed arrows represent inputs. (*) Both theory and data modules can have internal dependence of cosmological parameters, which are handled by a \clmm\ cosmology module (see Section~\ref{sec:cosmology}). (**) \clmm\ does have some functionality for parameter estimation (see Section~\ref{sec:samplers})
    .
    }
    \label{fig:pipeline}
\end{figure}

Motivated by the cosmological constraining power of galaxy cluster lensing, \clmm\ was envisioned as a flexible, evolving toolkit that could be used to overcome the challenges to exploiting that potential, benefiting \desc’s Clusters Working Group (CL WG) and the cluster lensing community at large. 
Traditionally, software for cluster lensing forecasts and data analysis have been separate, with (sometimes public) theory codes operating in the space of non-directly observable quantities (e.g., \code{cluster-toolkit} used in \citet{2019MNRAS.482.1352M}, \code{cluster-lensing} from \citet{2016AJ....152..228F}) and non-public data analysis codes in another space of directly observable quantities that may not be defined in high-level cosmological models. 
The net effects of this division are that there is not necessarily a guarantee of self-consistency of an end-to-end analysis and that an end-to-end pipeline may be opaque to outsiders who wish to confirm the results of an analysis; 
it is not generically possible to validate an entire pipeline with assumptions shared among all stages, particularly to outsiders seeking to reproduce the results should the data become public.
\clmm\ aims to unify the software pipeline infrastructure for cluster lensing to eliminate this division and enable complete self-consistency of future analyses within and beyond \desc.
In Figure~\ref{fig:pipeline} we present a pipeline diagram of \clmm's role in cluster weak lensing analyses.

In addition to this initial motivation, \clmm\ co-evolved with other goals targeting benefits whose relevance has risen since the initiation of the project.
\begin{itemize}
 \item \clmm's primary purpose is to serve as the \be\ for the \desc\ cluster lensing software pipeline, meaning it must be flexible enough to interface with other interconnected \desc\ codes that may themselves be considered moving targets and that it must be extremely well-tested and well-documented.
 \item \clmm\ is agnostic of the status of data as simulated or observed and includes tools connecting theory-motivated unobservable and data-oriented observable quantities, enabling robust end-to-end pipeline validation under the introduction of sources of systematic error and alternative prescriptions for the conversions.
 \item \clmm\ is a lightweight, modular toolkit rather than an end-to-end pipeline in and of itself, which maximizes its flexibility for use in a variety of analyses, liberates the development workflow from the overhead of wrapping whole optimization and sampling routines, and enables further extensions to be developed concurrently for specific science goals by sub-teams driven by a sense of ownership of the applications in question.
 \item \clmm\ supports several theory engines and includes extensive examples both of supporting code for interfacing with real data sets and of its usage in building pipelines presented as Jupyter Notebooks, thereby maximizing the potential user base as well as current and future capabilities of the code.
\end{itemize}

The \clmm\ team has endeavored to build and maintain a healthy software development environment by following best practices.
\clmm\ is open-source software developed publicly on GitHub.
We engage in test-driven development with continuous integration and code reviews to ensure sufficient documentation and unit test coverage at all stages.
We explicitly welcome new collaborators to join the team, which is comprised of theorists and observers, members and non-members of \desc, and those across all career stages, from undergraduates to senior faculty.
\clmm's v1.0 release encompasses the maximum functionality necessary before embarking on science-driven applications that correspond to synchronous development of the code in diverse directions, some of which are discussed in Section~\ref{sec:discussion}.


\section{Structure}
\label{sec:stucture}

\begin{figure*}
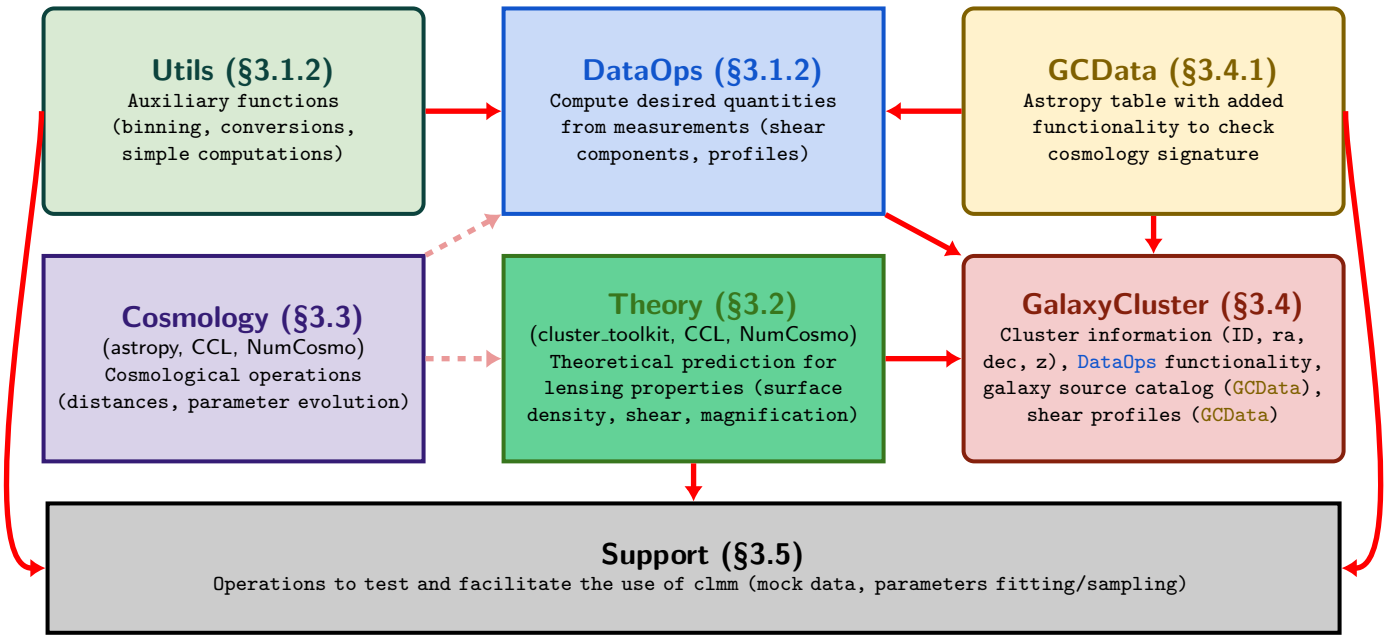

    \centering
    \diagram
    \caption{Diagram illustrating structure of code with dependencies (arrows).
    Sharp edge rectangles represent packages, rounded edge rectangles are modules, continuous arrows are direct imports, and dashed arrows mean the module is passed as an argument.
    }
    \label{fig:code_overview}
\end{figure*}

As a modular toolkit rather than an end-to-end pipeline itself, \clmm\ is intended to not only provide support for galaxy cluster lensing calculations performed by both theorists and observers but to do so in a way that does not permit the introduction of mutual inconsistencies that can invalidate an analysis built with \clmm. 
\clmm\ aims to prevent the subtle conflicts that may arise between assumptions in different parts of a complex analysis pipeline.
For example, \clmm\ enables users to prevent the inadvertent evaluation of a function of physical quantities derived from observables under different assumed cosmologies.

In Section~\ref{sec:dataops_module}, we first provide a quick introduction to the observables of weak lensing, before describing the data operations package that allows us to prepare the data vector from galaxy catalogs. 
The theory subpackage, described Section~\ref{sec:theory}, covers all the necessary computations regarding weak lensing quantities given a halo profile, mass definition and underlying cosmology model, the latter of which is described in Section~\ref{sec:cosmology}. 
In Section~\ref{sec:galaxycluster}, we present the \code{GalaxyCluster} object, the core data type of \clmm, as well as the motivations for its design.
Included in the base \clmm\ library are a set of useful functionalities, such as generating mock datasets, that are finally presented in Section~\ref{sec:support}. 
Figure~\ref{fig:code_overview} provides a schematic overview of the structure of the code and of the paper.

\paragraph*{A note on dependencies, units and constants} Before proceeding with the description of the code, we note that in addition to specific packages used for the theoretical predictions of weak lensing quantities (\S\ref{sec:modelbackends}) \clmm\ depends on the following standard python libraries: Astropy \citep{astropy:2018}, Matplotlib \citep{hunter_matplotlib:_2007}, Numpy \citep{harris_numpy_2020}, and SciPy \citep{virtanen_scipy_2020}.

We also clarify that \clmm\ uses physical units:
\begin{itemize}
    \item By default, all distances in \clmm\ are in Mpc, non-comoving.
    \item Cluster masses are in M$_\odot$.
  \end{itemize}
In short, we specify that units have no $h$-dependence\footnote{$h=H_0/(100$\, km\, s$^{-1}$\,Mpc$^{-1}$), with $H_0$ the Hubble constant at $z=0$.}.
A set of physical constants (e.g $G$, $c$, M$_\odot$) and unit conversion factors are also defined in \code{constants.py}. 
These come from CODATA 2018 \citep{codata18} or from the definitions of the IAU 2015 \citep{2015arXiv151007674M} when not available in the former.

\subsection{The \code{dataops} package}
    \label{sec:dataops_module}

The \code{dataops} package includes functions that may be used to construct profiles of radially averaged shear quantities from a sample of background galaxies.
We briefly introduce the weak lensing basics and the definitions of the weak lensing observables in Section~\ref{sec:datatheory}. 
In Section~\ref{sec:dataops}, we then present the operations needed to construct the weak lensing profiles and their correspondence to functions in the \code{dataops} subpackage and \code{utils} module. 
 
    \subsubsection{Physical definitions and equations of the weak lensing observables}
    \label{sec:datatheory}

Here we review the standard definitions and transformations used in weak lensing analyses that are implemented as functions in the \code{dataops} subpackage. We refer the reader to review articles such as  \citet{bartelmannandschneider2001} or \citet{2020A&ARv..28....7U} for a more thorough introduction.  

 \paragraph*{Lensing transformation}
 
The weak lensing distortion of the image of background sources is given by the Jacobian matrix $\cal{A}$ (also called the \emph{amplification} matrix), which maps the connection between the unlensed and lensed coordinate systems: 
\begin{equation} 
    \label{eq:A2}
 {\mathcal{A}}=(1 - \kappa )\left[{\begin{array}{c c }1- g_1&- g_2\\-g_2&1+g_1\end{array}}\right].
\end{equation}
Here $g=g_1+i g_2$ (in complex notation) is the reduced shear, the quantity measured in weak lensing analyses, and it is related to the gravitational convergence $\kappa$ and shear $\gamma$ through
\begin{equation}
    g_{1,2} =\frac{\gamma_{1,2}}{1-\kappa}.
    \label{eq:red_shear}
\end{equation}
The convergence and shear represent
the isotropic and anisotropic focusing of light rays, respectively. 
The subscript on $g$ and $\gamma$ refers to the direction of the gravitational distortion, where 1 refers to distortion along the axes in a Cartesian framework and 2 to a coordinate frame rotated by 45 degrees.

The determinant of $\cal{A}$ is related to the image magnification that describes the apparent change of flux of lensed background objects, 
     \begin{equation}
    \label{eq:magnification_from_A}
\mu = \frac{1}{\mathrm{det} \cal{A}} = \frac{1}{(1-\kappa)^2 - |\gamma|^2}\;.
     \end{equation}
     
The convergence and shear are observationally intertwined but in the regime of weak deformation, i.e. the \emph{weak lensing regime},  $\kappa\ll 1$ and $\gamma\ll1$ and we can linearly approximate the observables as : $\mu \approx 1 + 2\kappa$ and $g \approx \gamma$. 
Given the above definitions, the image of a circular source of radius $R$ is transformed into an ellipse with semi-axes $a$ and $b$ as: 
\begin{align*}  
 a = \frac{R}{1 - \kappa - |\gamma|} = \frac{R}{(1 - \kappa)(1 - |g|)}, \\
 b = \frac{R}{1 - \kappa + |\gamma|} = \frac{R}{(1 - \kappa)(1 + |g|)}.
\end{align*}  
If $|g|<1$, the axis ratio $q=b/a$ directly gives the absolute value of the reduced shear as $|g| = \frac{1-q}{1+q}$.

 \paragraph*{Measuring galaxy shapes}

Two common definitions of ellipticity  used in the literature \citep{great3} to describing galaxy shapes in lensing studies are:
 \begin{equation}
     |\epsilon| = \frac{1-q}{1+q},\, 
     |\chi| = \frac{1-q^2}{1+q^2}, 
 \end{equation}
 where $\epsilon = \epsilon_1+i\epsilon_2$ and $\chi = \chi_1 + i\chi_2$. 
Note that $|\epsilon|$ and $|\chi|$ are bound between 0 and 1. 
These two ellipticity definitions can be related through:
    \begin{equation}
    \label{eq:epsilon_to_chi}
      \epsilon = \frac{\chi}{1+(1-|\chi|^2)^{1/2}}\;,\quad
      \chi = \frac{2\epsilon}{1+|\epsilon|^2}\;.
    \end{equation} 
In the remainder of this section, the discussion focuses on the $\epsilon$ ellipticity. 
For $|g|\le1$, it can be shown that the transformation between the true intrinsic ellipticity of a source $\epsilon^{\rm (true)}$ and the lensed ellipticity of its image $\epsilon^{\rm (lensed)}$ is given by 
\begin{equation}
\label{eq:transform}
  \epsilon^{\rm (lensed)} = \frac{\epsilon^{\rm (true)}+g}{1+g^*\epsilon^{\rm (true)}}\;.
\end{equation}

As the intrinsic shape of galaxies is unknown, the key assumption needed for weak lensing analyses is that galaxy orientations are random, and thus that for a large enough sample the ellipticities average out to zero: $\langle \epsilon^{\rm (true)}\rangle \approx 0 $. 
For a region in which the shear is approximately constant, it then follows from Eq. \ref{eq:transform} that, in the weak lensing regime, where $|g|\ll1$, 
    \begin{equation}
    \label{eq:epsilon_avg}
\langle \epsilon^{\rm (lensed)} \rangle = \left<\frac{\epsilon^{\rm (true)}+g}{1+g^*\epsilon^{\rm (true)}}\right> \approx  \left<\epsilon^{\rm (true)}+g\right> \approx g .
    \end{equation}   
Therefore, by averaging the shapes of the lensed galaxy images, we can locally estimate the reduced shear. 
We note that a common source of bias in the shear estimate stems from contamination of the lensed source galaxy sample with unlensed cluster members. 
This current version of \clmm\ does not implement functionalities to compute the boost factors used to correct for this effect (e.g., \citealp{applegate14,2017MNRAS.469.4899M}); this is left for future work.

 \paragraph*{Tangential and cross components}
The standard simplification of a galaxy cluster is the assumption of a circularly-symmetric matter distribution. We note that this is a only reasonable on average for a large sample of clusters, as individual cluster mass distributions are triaxial and have substructures. 
For an ideal symmetric cluster the shear is only oriented tangentially to the radial direction due to symmetry. In this case polar coordinates are easy to work with.

The two components of the shear, reduced shear, and ellipticities are defined relative to a chosen Cartesian reference point, usually the lens center.
Using the position angle $\phi$ of the source relative to the lens center, we can  decompose the complex quantities in a \emph{tangential} and \emph{cross} terms. For the reduced shear this gives, 
\begin{equation} 
    g_{+}= - g_1\cos\left(2\phi\right) - g_2\sin\left(2\phi\right)\;,
\label{eq:shear_t}      
\end{equation}   
and
\begin{equation}       
g_\times = g_1 \sin\left(2\phi\right) - g_2\cos\left(2\phi\right)\;,
\label{eq:shear_x}      
\end{equation}
where $g_{+}$ is the reduced tangential shear and $g_\times$ is the reduced cross shear.
Hence, 
\begin{equation}
    \label{eq:shear} \nonumber
    g_{+} = |g|\;  \quad\mathrm{and}\quad  g_\times = 0 \;.
\end{equation}  
In practice, because the cross-component of the reduced shear is expected to be null for an ideally spherical mass distribution, it provides a useful test for systematic uncertainties in the shear measurement. 

    \subsubsection{Data operations}
    \label{sec:dataops}

\begin{table*}
\centering
\begin{tabular}{ l p{3in} c}
 \hline
    Function & Short description$^\dagger$  & Equations \\
 \hline
 \hline
 \code{compute\_lensed\_ellipticity}\dotfill & compute the lensed ellipticities from intrinsic ellipticities, $g$ and $\kappa$ & \eqref{eq:transform}\\
 \code{compute\_radial\_average}\dotfill & return the arithmetic mean and corresponding statistical errors in each bin & \eqref{eq:radial_average}  \\
\code{convert\_shapes\_to\_epsilon}\dotfill & convert between ellipticity/shape definitions & \eqref{eq:epsilon_to_chi}\\
 \code{convert\_units}\dotfill & convert between different angular and physical distance units & -- \\
 \code{make\_bins}\dotfill  & return an array of radial bin edges, given user-defined options  & -- \\
 \hline
 \code{compute\_tangential\_and\_cross\_components}\dotfill & compute the angular separation to the source center and the tangential and cross values given shape parameters & \eqref{eq:shear_t} and \eqref{eq:shear_x}\\
  \code{make\_radial\_profile}\dotfill & return the average values of the tangential and cross components and associated errors in each bin   & \eqref{eq:radial_average} \\
 \hline
\end{tabular}
\caption{
Summary table of the main functions implemented either in \code{utils.py} (top) or \code{dataops.py} (bottom), that are used to perform or support data operations. 
$^\dagger$ The descriptions correspond to the main way to use these functions, though more options and functionalities, described below, are available to the user.
\label{tab:dataops}}
\end{table*}

A pipeline constructing galaxy cluster weak lensing profiles using \clmm\ may be built from two sets of operations that constitute the core of the \clmm\ \code{dataops} package. 
This module is itself based on several helper functions, also directly available to the users, hosted in the \code{utils} module. 
Those two sets of operations are referred to as ``data operation modules'' in the following and Table~\ref{tab:dataops} summarizes the functions of those modules.
 
As seen in Section \ref{sec:datatheory}, in observed data sets, one can access the shear quantities from the measurements of the source galaxy shapes. 
Alternatively, one can access intrinsic and cosmological quantities from a simulated data set. 
The data operation modules allows for the use of these different kinds of quantities as input.

\paragraph* {The \code{dataops} package} contains tools to compute radially averaged profiles of shear quantities around lenses. 
It is based on two sets of operations.
\label{sec:dataops_ops}

First, \code{compute\_tangential\_and\_cross\_components} takes as input a galaxy catalog with coordinates and the two components of a complex weak lensing quantity, and returns, for each galaxy, the angular separation $\theta$ with respect to the cluster center\footnote{The default is to compute the angular separation using the method from \code{astropy.Coordinates} that performs the full computation on the sphere. The flat-sky approximation, that will generally hold on typical cluster field size, is also an option of the code.}, and the tangential and cross components of the corresponding input lensing quantity following Eqs. \ref{eq:shear_t} and \ref{eq:shear_x}.
The input components $g_1$ and $g_2$ can correspond to the complex ellipticity of the source galaxies (as can be measured in observed data sets), but also to the complex shear $\gamma$ or reduced shear $g$ (that can be accessible from e.g. a simulated data set). 
If set by the user, the input components are multiplied by the critical surface mass density $\Sigma_{\rm crit}$ (that depends on the distances of the observer-cluster-source system according to Eq.~\eqref{eq:crit_dens}). 
The latter can be either given by the user, or computed from the theory module if a cosmology object and the redshifts of the lens and sources are also provided (see Section \ref{sec:theory}). 
This operation can transform the shear $\gamma$ into the excess surface density $\Delta \Sigma$ (a quantity directly related to the density distribution in the cluster according to Eq.~\eqref{eq:tang_shear}), or $g$ and $\epsilon$ into estimates of $\Delta \Sigma$, as generally done in the literature.

Second, \code{make\_radial\_profile} allows the binning of the shear quantities as a function of angular separation $\theta$ to return the mean tangential and cross shear profiles and their corresponding statistical errors. 
Optionally, the list of galaxies in each bin may also be returned. 
The user may provide a list of bin edges, which \clmm\ provides as an optional helper function, \code{make\_bins} (see \S\ref{sec:utils}). 
Profiles may be computed in angular or physical units corresponding to a given cosmology. 
The majority of the functions enabling \code{make\_radial\_profile} are stored in the \code{utils} module and may be accessed directly by the user. 

\paragraph*{The \code{utils} module}
\label{sec:utils} 
includes several helper functions to facilitate the construction of radially averaged profiles and the conversion of quantities from both observed and simulated data-sets:
\begin{itemize}
    \item \code{compute\_lensed\_ellipticity} computes the lensed ellipticities from the intrinsic ellipticities, shear, and convergence, following Eq. \ref{eq:transform}.
    \item \code{convert\_shapes\_to\_epsilon} converts $\chi$ to $\epsilon$ ellipticity definition, following Eq. \ref{eq:epsilon_to_chi}, or converts shear $\gamma$ into reduced shear $g$, given a convergence value, following Eq. \ref{eq:red_shear}.
    \item  \code{make\_bins} returns radial bin edges given a minimum and a maximum limit and a number of bins, via the user's specification of even spacing, log-spacing, or evenly populated bins.
    \item \code{compute\_radial\_average} bins $y$ values into bins, based on the specified bin edges, and currently computes for each bin $i$ the average $\langle y \rangle_i$ and associated statistical errors $\sigma_i$ of all $y$ simply as
    \begin{equation}
        \langle y \rangle_i = \frac{1}{N_i}\sum_{j=0}^{N_i} y_j\;,\; \sigma_i=\frac{{\rm stddev}_i}{\sqrt{N_i}}\;, 
        \label{eq:radial_average}
    \end{equation}
    where $N_i$ and ${\rm stddev}_i$ are the number of galaxies and the standard deviation of the values of $y$ in bin $i$, respectively. 
    The number of objects in each bin and the list of bin indices for each $y$ entry are also returned. 
    Future releases of \clmm\ will incorporate shape measurement errors into the total error estimate, support non-Gaussian error distributions, and accommodate user-provided weights, enabling, e.g. the optimal weighting scheme of the weak lensing $\Delta\Sigma$ estimator \citep{2004AJ....127.2544S}.
    \item  \code{convert\_units} facilitates conversion between different angular and physical units, given a cosmology object as defined in Section \ref{sec:cosmology}.
\end{itemize}

\subsection{The \code{theory} package}
\label{sec:theory}

\clmm\ isolates all functionality that requires an assumed cosmological model in the \code{theory} subpackage.  
The reasoning behind this is to enforce consistent assumptions in multiple parts of a \clmm\ analysis, which is of critical importance given the variety of analysis pipelines that could be built from the otherwise shared functionality within \clmm. 

In Section~\ref{sec:math} we present the most relevant equations regarding the halo profiles and weak lensing functions.  
\clmm\ relies on the public codes described in Section~\ref{sec:modelbackends} to provide models of mass density profiles and gravitational lensing quantities; 
these have been thoroughly cross-checked using a validation procedure described in Section~\ref{sec:validation}.
Finally, we describe in Section~\ref{sec:implementation} the functional and object-oriented implementation structures and also provide the names of the methods and functions. 
\subsubsection{Physical definitions and equations: Modeling mass distribution and weak lensing quantities} 
\label{sec:math}

The ultimate task of \clmm\ is to provide tools to enable users to estimate cluster masses from lensing observables, primarily 
the reduced shear $g_\mathrm{t}$ \citep{hoekstra03}.  
To estimate cluster masses from these observables, we first parameterize the three dimensional mass distribution of a galaxy cluster and model the projection effects on weak lensing observables.

One way to characterize the structure of dark matter halos is to calculate the spherically averaged mass density profile $\rho(r)$, where $r$ is the distance to the cluster center.  
We then parametrize the mass profile based on a given {\it model profile}.  
Though \clmm\ currently supports only spherical galaxy cluster shapes, future code development will account for the triaxial shape of clusters. 
\clmm\ supports the following model profiles from the \be s described in Section~\ref{sec:modelbackends}: 
\begin{itemize}
    \item  The two-parameter profile 
    \begin{equation}
    \label{eq:rho_NFW_Hernquist}
      \rho(r) = \frac{\rho_\mathrm{s}}{\left(r/r_\mathrm{s}\right) \left(1 + r/r_\mathrm{s}\right)^n},
    \end{equation}
    which is the Navarro-Frenk-White \citep[NFW;][]{NFW1997} profile for $n=2$, and the \citet{hernquist1990} profile for $n=3$.
    \item The three-parameter profile, \citet{einasto1965},
    \begin{equation}
    \label{eq:rho_einasto}
        \rho(r) = \rho_\mathrm{s} \exp{\left\{ - \frac{2}{\alpha} \left[ \left(\frac{r}{r_\mathrm{s}}\right)^\alpha - 1 \right] \right\} },
    \end{equation}
    where $\alpha$ is a shape parameter.
\end{itemize}

Both the scale radius $r_\mathrm{s}$ and the characteristic density $\rho_\mathrm{s}$ are functions of the cluster's mass $M_\Delta$ and the concentration $c_\Delta$. 
Under the spherical overdensity mass definition used in this first release of \clmm, 
\begin{equation}
    r_\mathrm{s} = \frac{1}{c_\Delta} \left( \frac{3 M_\Delta}{4 \pi \Delta \rho_\mathrm{bg}} \right)^{1/3},
\end{equation}
where $\rho_\mathrm{bg}$ is the background matter density (mean or critical) and $\Delta$ is the overdensity with respect to the background density (e.g., 200, 500, or the overdensity at the virial radius of the cluster $\Delta_{\mathrm{vir}}$ given by the approximation of \cite{Bryan1998}). 
The characteristic density is
\begin{equation}
    \rho_s = \frac{c_\Delta^3 \Delta \rho_\mathrm{bg}}{3 I(c_\Delta)},
\end{equation}
where 
\newcommand{\de}[1]{\left(#1\right)}
\begin{equation}
    I \equiv \int_0^{c_\Delta} \mathrm{d}x\; x^2 \hat{\rho}(x),
\end{equation}
where $x=r/r_s$ and $\hat{\rho}(x)\equiv\rho(xr_s)/\rho_s$ is the dimensionless density.

The surface mass density is obtained by integrating the 3D profile $\rho(r)$ along the line of sight as
\begin{equation}\label{eq:Sigma}
    \Sigma (R) \equiv \int_{-\infty}^{\infty}  \mathrm{d}y \, \rho(r,y),
\end{equation}
where $R$ is the projected radial distance from the center of the cluster.
The average surface mass density within a disk of radius $R$ is then
\begin{equation}\label{eq:mean_Sigma}
    \overline{\Sigma} (<R) = \frac{2}{R^2} \int_0^R \mathrm{d}R^\prime \, R^\prime \Sigma(R^\prime).
\end{equation}

The convergence $\kappa(R)$ is defined as the dimensionless surface mass density
\begin{equation}
    \label{eq:convergence}
    \kappa(R) = \frac{\Sigma(R)}{\Sigma_\mathrm{crit}},
\end{equation}
where $\Sigma_\mathrm{crit}$ is the critical surface mass density
\begin{equation}
\label{eq:crit_dens}
    \Sigma_\mathrm{crit} = \frac{c^2}{4 \pi G} \frac{D_\mathrm{S}}{D_\mathrm{L} D_\mathrm{LS}},
\end{equation}
and $c$ is the speed of light, $G$ is the gravitational constant, $D_\mathrm{S}$, $D_\mathrm{L}$ are the angular diameter distances from the observer to the source (lensed background galaxy) and lens (cluster), respectively, and $D_\mathrm{LS}$ is the distance between lens and source. 

The tangential shear is 
\begin{equation}
    \label{eq:tang_shear}
    \gamma_t(R) = \frac{\overline{\Sigma} (<R) - \Sigma (R)}{\Sigma_\mathrm{crit}} =  \frac{\Delta\Sigma(R)}{\Sigma_\mathrm{crit}} ,
\end{equation}
where $\Delta\Sigma(R)$ is the excess surface density at projected radius $R$.
Using expressions of the shear and convergence above, the predicted reduced shear $g_t(R)$ and the magnification $\mu(R)$ profiles are then directly obtained from Eqs~\eqref{eq:red_shear} and \eqref{eq:magnification_from_A}.

Note that for the NFW parametrization, analytical solutions exist for $\gamma$ and $\kappa$ as detailed in \citet{wright00}. 
The lensing quantities of Einasto and Hernquist require analytical integration (see discussion below). 

\begin{table*}
\centering
\begin{tabular}{ l l l l }
 \hline\hline
        & \code{cluster-toolkit [v1.0]} & \code{CCL [>v2.0]} & \code{NumCosmo [>v0.15]}\\
 \hline
 Halo profile\dotfill     & NFW    & NFW, \color{gray}{Einasto$^{\dagger}$, Hernquist$^\dagger$} & NFW, Einasto, Hernquist\\
 Mass definition\dotfill  & mean  & mean, critical, virial & mean, critical, virial\\
 Cosmology object\dotfill & \astropy cosmology & \ccl\ cosmology &  \nc\ cosmology\\
 \hline
\end{tabular}

$^\dagger$ The Einasto and Hernquist profiles from \ccl will soon become available in \clmm.
\caption{Properties of the three modeling \be s as available in CLMM, regarding the halo profile parameterization, halo mass definition and type of cosmology object. The \be\ versions required in \clmm\ are given with their names.}  \label{tab:backends}
\end{table*}

\subsubsection{Modeling \be s}\label{sec:modelbackends}

In this first release, \clmm\ offers the user three possible \be s to model the cluster weak lensing quantities described in \S\ref{sec:math}, specified by the \code{CLMM\_MODELING\_BACKEND} environment variable. 
Each \be, summarized in in Table~\ref{tab:backends}, relies on a different underlying cosmology object, each of which is wrapped as a subclass of the \clmm\ cosmology superclass described in Section~\ref{sec:cosmology}.

\paragraph*{\code{cluster-toolkit} (\ct)}\footnote{\url{https://cluster-toolkit.readthedocs.io/en/latest/index.html}} 
deals specifically with cluster lensing and has been used as part of the cluster analysis of \citet{2019MNRAS.482.1352M}. 
It provides all relevant computations through a functional interface but is limited to using overdensity masses defined with respect to the mean density of the universe $M_{\Delta,m}$.  
A user specifying the \ct\ \be\ may only access the NFW halo profile model, for which the cluster lensing quantities have analytical expressions.
The native \ct\ units ($h^{-1} M_\odot$ and $h^{-1}$Mpc) are converted to the physical units we use throughout \clmm.
As \ct\ does not provide its own cosmology object, \clmm\ employs the \astropy cosmology object for the \ct\ \be. 

\paragraph*{\ccl\ (Core Cosmology Library)}\footnote{\url{https://github.com/LSSTDESC/CCL}} 
is a library developed as a \desc\ tool to accurately compute cosmological observables \citep{2019ApJS..242....2C}. 
In contrast to \ct, it is not cluster-specific and provides predictions for many cosmological quantities (e.g., distances, correlation functions, power spectra), including but not limited to quantities relevant to cluster lensing. 
Using \ccl\ as the modeling \be\ of \clmm, the user may use the various standard overdensity mass definitions found in the literature, i.e., mean, critical and virial. 
So far, \clmm\ only supports \ccl's NFW halo parametrisation.   
Comoving distances are used in \ccl\ and appropriate conversion are implemented to meet the \clmm\ choice of physical units. 
\ccl\ is object-oriented and comes with its own cosmology class that is used in \clmm\ when \ccl\ is the selected \be.

\paragraph*{\code{NumCosmo} (Numerical Cosmology Library; \nc)}\footnote{\url{https://numcosmo.github.io/}}  \citep{2014ascl.soft08013D} 
is also a general object-oriented cosmology library, with functionalities extending beyond cluster-related quantities that also provides a comprehensive set of tools to analyze statistical models. 
Using this \be, the user may choose between the NFW, Einasto, or Hernquist halo parametrisations and use the mean, critical, or virial overdensity mass definition. 
\nc\ natively uses the same units as \clmm. 
If selected as \be, a \nc\ cosmology object is used to describe the cosmology. 
In this first release of \clmm, this is the \be\ that provides the most functionalities.

Depending on their goals and needs, the users may therefore choose any of these \be s, from the relatively straightforward \ct, to the more involved \ccl\ and \nc\ options. 
If the user does not specify the \code{CLMM\_MODELING\_BACKEND} environment variable, \clmm\ will default to the \ccl\ \be\ if it is installed and will otherwise  fall back on \nc\ and \ct.

\subsubsection{Back-end comparison and validation tests} \label{sec:validation}

In addition to the unit test coverage of \clmm, we apply a rigorous validation procedure with a focus on the theory \be s.
In this section, we justify our choices in theory \be\ support for each profile model presented in Table~\ref{tab:backends}.  We compare the per-\be\ predictions of the NFW, Einasto, and Hernquist profiles under controlled experimental conditions to inform our choices.  The validation tests share a common configuration: mass definition of $M_{200,\rm{m}}$, radial range of $10^{-3}-10$\;Mpc, cluster mass of $10^{15}$\;M$_\odot$, cluster redshift of $z=0.2$, and concentration $c=5$.
We also apply the appropriate conversions between the slightly different per-\be\ values of physical constants (e.g., speed of light or Newton's gravitational constant).
The notebooks to reproduce the results of these validation tests are available in the code's GitHub repository.

\begin{figure}
  \centering
  \includegraphics{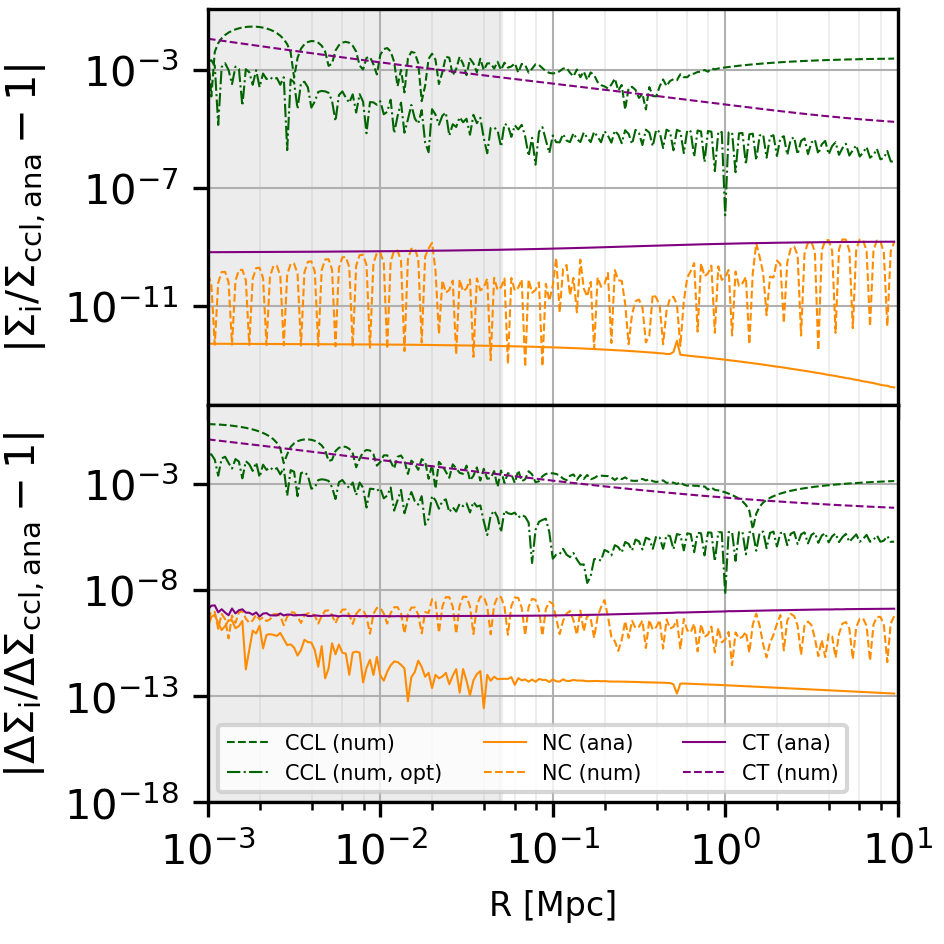}
  \caption{NFW Surface (top) and excess (bottom) surface mass density relative to the analytical \ccl\ implementation. Solid lines correspond to analytical implementations in \ct\ (purple) and \code{NC} (orange). Dashed lines show the results of the numerical implementations for \ct, \code{NC} and \ccl\ (green). The innermost regions shaded in grey are not used in typical cluster weak lensing analyses.}
  \label{fig:NFW_comparison} 
\end{figure}

\paragraph*{Mass and excess surface mass densities} 
All three \be s implement the analytic solutions for $\Sigma$ and $\Delta\Sigma$ with the NFW parameterization. We choose the analytic computation from \ccl\ as a reference for relative comparisons.
Figure~\ref{fig:NFW_comparison} shows the excellent agreement with the analytic implementations of both \ct\ (solid purple line) and \nc\ (solid black).  Here, the relative discrepancies in $\Sigma$ and $\Delta\Sigma$ never exceed $10^{-9}$ over the full radial range\footnote{\code{CT} does not use the same definition of the physical constants as \code{CCL} and \code{NC}, and some conversion is implemented in \clmm. Small rounding errors in the conversion are likely the source of the difference between the \code{NC} and \code{CT} analytical curves.}. From the software point of view, it is relevant to compare the results over the a wide radial range, but we note that regions below $R\sim 0.1$~Mpc are generally not used in most cluster weak lensing analyses.
Because the NFW parametrization has an analytic solution, this parametrization also allows us to test the performance of each \be\ when numerically computing $\Sigma$ and $\Delta\Sigma$.

For both quantities, we find \nc's numerical implementation to best match the precision of the analytic implementations, while \ccl\ and \ct\ (green and purple lines respectively) provide less accurate results, especially at small radii.
\ct does not provide user-defined precision parameters and requires users to tabulate the 3D density prior to obtaining $\Sigma$ and in turn $\Delta\Sigma$ at specified radii, necessitating checks to ensure consistency. The \ccl profile integration scheme relies on FFTLog\footnote{\url{https://jila.colorado.edu/\~ajsh/FFTLog/}} and is quite sensitive to the radial range and sampling used. The default out-of-the-box behaviour (green dashed line) can be improved by one to two orders of magnitude by changing some of the integration options and parameters (green dash-dotted line). Only the NFW analytical profiles are implemented in \clmm; the numerical solutions were used only to assess the reliability of the numerical integration schemes from the three backends.

\begin{figure}
  \centering
  \includegraphics{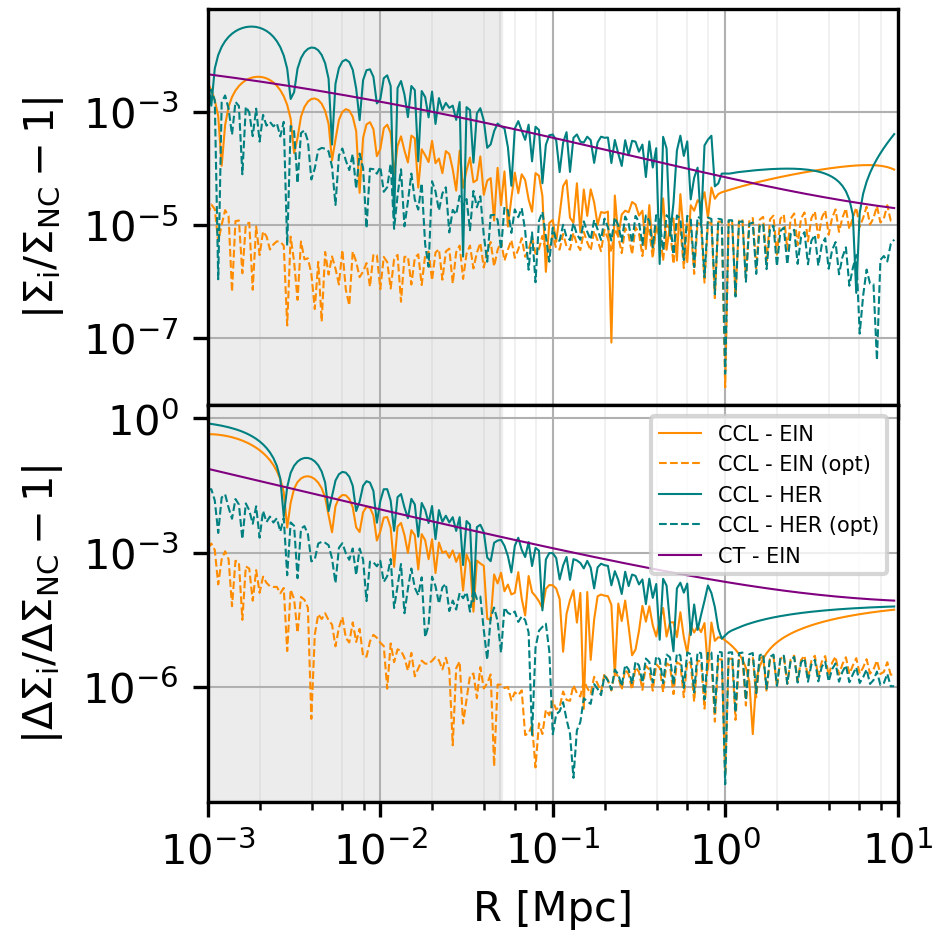}
  \caption{Same as Figure~\ref{fig:NFW_comparison} but for the Einasto (\ccl orange and \ct purple) and Hernquist profiles (\ccl blue), choosing the corresponding \nc result as reference. The Hernquist profile is not implemented in \ct. For \ccl, both the default (solid) or optimized (dashed) integrations are shown.}
  \label{fig:Einasto_comparison} 
\end{figure}

The Einasto and Hernquist profiles require a numerical integration. Figure~\ref{fig:Einasto_comparison} summarizes the relative precision of each \be\ relative to the \nc\ computation, that serves as benchmark given the high precision reached for the NFW case. Looking at \ct and the default setup for \ccl (solid lines), the absolute relative difference with \nc\ reaches the $\sim  1-10\%$ level at the innermost radii. For both the \ccl Einasto (orange) and Hernquist (blue) profiles, switching from the default (solid) to the optimized integration setup (dashed, defined from the NFW analysis) improves the agreement with the benchmark by several orders of magnitudes.

The results above have been obtained for a flat $\Lambda$CDM cosmology with no massive neutrinos. 
For the analytical NFW, we find that by adding one massive neutrino to the cosmology, the disagreement between \ccl\ and \nc\ remains at the $10^{-6}$ level over the full radial range. 
However it rises to the $10^{-2} - 10^{-3}$ level for \ct\ due to the fact that the \astropy cosmology supporting the \ct\ \be\ does not take into account massive neutrinos to compute the matter density parameter $\Omega_m$.

\paragraph*{Impact on the reconstructed mass} 

Any differences in the predicted profiles will impact the reconstructed mass. To test this effect, we choose 100 idealized data points from the analytic \ccl\ NFW and \code{NC} Einasto $\Delta\Sigma$ models. We then adjust the numerical \ct\ and default numerical \ccl\ NFW models to the idealized dataset obtained from the analytical \ccl\ model over the $[10^{\{-3,-1\}}-10]$\;Mpc range. 
We find that the reconstructed mass using the numerical predictions of \ccl\ and \ct\ are biased by $\sim 10\%$ and $\sim 6\%$, respectively, when fitting over the $[10^{{-3}}-10]$\;Mpc range. 
Using instead the range of $[10^{{-1}}-10]$\;Mpc, typically used in cluster WL analyses, the biases in the reconstructed mass are reduced to 0.3\% and 0.1\% respectively; 
this is subdominant in the mass estimation error budget of a fit on real data subject to the list of systematic effects identified in Section~\ref{sec:intro}. For \ccl, this could be improved further using the optimized FFTLog integration setup mentioned previously.
Repeating the same exercise for the numerical \nc\ NFW profile yields an unbiased mass estimation (bias $<10^{-5}\%$ whatever the radial range). 

In light of this, we only include in this first release of \clmm the Einasto and Hernquist pofiles from \nc\ because of the more accurate performance of the default, out-of-the-box, \nc integration. We note however the good performance that can also be reached ($\sim 10^{-5}$ level) when tuning the FFTLog integration of \ccl; this needs to be explored further before being implemented robustly in \clmm, given the sensitivity of the results to the configuration choices (e.g., radial range, sampling). Finally, an unreleased version of \ccl, implementing an integration in real space, showed that precision could be increased even further. In light of these near-future developments, we anticipate that the Einasto and Hernquist parametrisations from \ccl will soon be implemented in \clmm.

\subsubsection{Implementation}
\label{sec:implementation}

\clmm\ provides both a functional and an object-oriented interface. 
Users of the functional interface provide input parameters to functions that perform computations each time the function is called, which is convenient when a function is called only a few times, as may be anticipated in data exploration and pedagogical applications.
Users may also access the object-oriented interface for the \ccl\ and \nc\ \be s, which differs in that users may access quantities previously computed by class methods, e.g. a galaxy cluster's halo profile or the cosmological model used for multiple analysis stages, as attributes of an object. 
This ensures that the methods computing the weak lensing quantities (see Table~\ref{tab:function_names}) can be called multiple times without any overhead and that consistent assumptions are used between pipeline stages built using \clmm. 
To ensure API consistency between all \be s, there is also an object-oriented layer around the functional \ct\ interface, however, it has not been optimized.   

\begin{table*}
\centering
\begin{tabular}{  l c c c   }
 \hline \hline
    Function  & Symbol  & Equations & Units \\
 \hline
 \code{*\_3d\_density}\dotfill     & $\rho(r)$    & \eqref{eq:rho_NFW_Hernquist} and  \eqref{eq:rho_einasto}\dotfill & M$_\odot$\;Mpc$^{-3}$ \\
 \code{*\_surface\_density}\dotfill & $\Sigma(R)$ & \eqref{eq:Sigma} &  M$_\odot$\;Mpc$^{-2}$ \\
 \code{*\_convergence}\dotfill & $\kappa(R)$ & \eqref{eq:convergence} & \textemdash \\
 \code{*\_critical\_surface\_density}\dotfill  & $\Sigma_{\rm crit}$  & \eqref{eq:crit_dens} & M$_\odot$\;Mpc$^{-2}$ \\
 \code{*\_excess\_surface\_density}\dotfill & $\Delta\Sigma(R)$ & \eqref{eq:tang_shear} & M$_\odot$\;Mpc$^{-2}$ \\
 \code{*\_tangential\_shear}\dotfill & $\gamma_t(R)$ & \eqref{eq:tang_shear} & \textemdash \\
 \code{*\_reduced\_tangential\_shear}\dotfill & $g_t(R)$ & \eqref{eq:red_shear} and \eqref{eq:tang_shear} & \textemdash \\
 \code{*\_magnification}\dotfill & $\mu(R)$ & \eqref{eq:magnification_from_A},\eqref{eq:convergence},\eqref{eq:tang_shear} & \textemdash \\
 \hline
\end{tabular}
\caption{Functions implemented in the functional and object-oriented approaches, where * reads as \code{compute} or \code{eval}, depending on the use of the theory functional or object-oriented interface. We also indicate the respective equations, defined in Section~\ref{sec:math}, and the units of each physical quantity.\label{tab:function_names}}
\end{table*}

The names of the functions and methods in both approaches differ only in the prefix of \code{compute\_*} for functions and \code{eval\_*} for class methods. 
Table~\ref{tab:function_names} summarizes the functions/methods implemented. 

\subsection{The \code{CLMMCosmology} objects}
\label{sec:cosmology}

As described in Table~\ref{tab:backends}, each theory \be\ relies on a different cosmology object. 
\astropy cosmology support has been set up for \ct, while we naturally kept \ccl\ and \nc\ native cosmology objects in these two latter cases. 
To make this seamless for the user, a \clmm\ cosmology superclass has been implemented to support the various \be\ cosmology objects. 
The \code{CLMMCosmology} superclass, defined in \code{clmm\_cosmo.py}, allows the user to set the values of the following cosmological parameters: 
the Hubble constant and the present-day dark matter, baryon and curvature density parameters, respectively, $H_0$, $\Omega_{c,0}$, $\Omega_{b,0}$ and $\Omega_{k,0}$. 
The cosmology class only hosts the few methods relevant to cluster WL mass reconstruction and that depend only on the cosmological parameters and redshifts. 
These include angular diameter distances, critical surface density and angular to physical unit conversion.

The default background model is flat $\Lambda$CDM, and the default values of the cosmological parameters are the Planck18+BAO best-fit estimates \citep{planckcosmology18}\footnote{See the sixth column of Table 2 in \cite{planckcosmology18}}, i.e., $H_0 = 67.66$, $\Omega_{c,0} = 0.262$, $\Omega_{b,0} = 0.049$ and $\Omega_{k,0} = 0.0$. 
The other parameters are $\sigma_8 = 0.8102$, spectral index $n_s = 0.9665$, CMB temperature today $T_{\gamma, 0} = 2.7255 \, \text{K}$, effective neutrino number $N_{{\rm eff}} = 3.046$ and one massive neutrino $m_\nu = 0.06 \; \text{eV}$ (and two massless).
This implementation was designed to facilitate the usage when there is no need to change this background cosmological model. 
To alter the background model, i.e., modifications on any parameter but $H_0$, $\Omega_{c,0}$, $\Omega_{b,0}$ and $\Omega_{k,0}$, the user must initialize the cosmology object of the specific modeling \be\ and then pass the instantiated object to the \clmm cosmology superclass.

\subsection{The \code{GalaxyCluster} class}
    \label{sec:galaxycluster}

The \code{GalaxyCluster} object is the core data structure of \clmm\ that stores relevant cluster weak lensing quantities and provides ready interfacing with \code{dataops} functions.

\subsubsection{Attributes}

A \code{GalaxyCluster} object corresponds to a single cluster with the following attributes commonly used in cosmology calculations: 
unique string or integer identifier \code{unique\_id}, right ascension and declination of the cluster center in degrees (\code{ra\_cl}, \code{dec\_cl}), redshift of the galaxy cluster \code{z\_cl}, and \code{GCData} object containing a catalog of (background) galaxy properties to be used in the weak lensing analysis \code{galcat}.

These quantities come with some caveats. 
For example, the center of a galaxy cluster is an ill-defined quantity as it is not directly observable, commonly but not universally estimated as the position of the brightest cluster galaxy in the optical; yet it is an essential variable in shear profile calculations. 
In order to avoid introducing inconsistent assumptions into an analysis pipeline built with \clmm, users considering multiple values of a single physical galaxy cluster's properties, such as center definitions or redshift estimates, must instantiate separate \code{GalaxyCluster} objects for each set of attributes.

\paragraph*{The \code{galcat} attribute} 
\label{sec:gcdata}
is a \clmm\ \code{GCData} object, which for this release of \clmm, is an \astropy \code{Table}, with added functionality to tag a cosmology (when the data itself assumed a specific cosmology) and a modified \code{print} function. This table includes support for a variety of data fields.  At minimum, \code{galcat} should contain unique identifier (\code{galaxy\_id}), sky coordinates in degrees (\code{ra}, \code{dec}), redshift (\code{z}), and two shape components (e.g., ellipticity components $\epsilon_1$ and $\epsilon_2$) for each background galaxy entry. Default \astropy \code{Column} names for the shape components are \code{e1} and \code{e2}, but the user may choose other names. This table may optionally contain other columns. 
Example optional columns include, shear components and convergence that may be directly measured in simulated data, or the redshift probability density function of each galaxy that will be required in most analyses. \code{GalaxyCluster} methods will also supplement this table with new fields as described below.

\paragraph*{A note on ensuring consistency}: 
Safeguards have been implemented on the \code{GCData} class to ensure the consistency of the assumed cosmology between operations.
If a cosmology is required to compute something in the \code{gc.galcat} or \code{gc.profile} (e.g., to compute $\Sigma_{\rm crit}$ or compute the binned profile as a function of the projected physical distance), the cosmology object is saved as metadata of the \code{GCData} table. 
Subsequently, an error will be raised if a different cosmology object is provided to a subsequent operation requesting a cosmology, thereby preventing users from accidentally introducing inconsistencies into an analysis built with \clmm.

    \subsubsection{Methods of the \code{GalaxyCluster} object}
    \label{sec:gcmethods}
\clmm\ provides access to the following class methods of an instance of the \code{GalaxyCluster} object \code{gc}:

\begin{itemize}
    \item \code{add\_critical\_surface\_density} adds a \code{sigma\_c} column to the \code{galcat} table, computed according to Eq.~\ref{eq:crit_dens} given a cosmology and using the cluster redshift \code{gc.z} and galaxy redshifts \code{gc.galcat['z']}. 
    It also tags the assumed cosmology used into \code{galcat}.

    \item \code{compute\_tangential\_and\_cross\_components} returns the angular separation from the cluster center in radians and the tangential and cross components according to Eqs.~\ref{eq:shear_t} and \ref{eq:shear_x} using the corresponding function of \code{dataops.py} (Section~\ref{sec:dataops}). 
    If not otherwise specified by the user, the operation will be performed on the columns named \code{'e1'} and \code{'e2'} of the \code{galcat} table. 
    If the \code{is\_deltasigma} keyword is set to \code{True}, the tangential and cross components will be multiplied by the critical surface density method to provide an estimate of $\Delta\Sigma$. 
    If the user sets the \code{add\_to\_cluster} keyword to \code{True}, the results are also added as new columns of the \code{galcat} table. 
    The angular separation is stored in \code{gc.galcat['theta']}, while the user is free to choose the name for the tangential and cross components columns (with defaults of \code{'et'} and \code{'ex'}). 

    \item \code{make\_radial\_profile} uses the corresponding function of \code{dataops.py} to return the azimuthally-averaged quantities in radial bins (user-defined bin units) from the cluster center. 
    The default is to use the \code{'et'} and \code{'ex'} fields, but the user can specify what fields to use from the \code{galcat} table. 
    By default, the resulting table is stored in a new attribute of the \code{GalaxyCluster} object, named \code{profile}, though the user has the option to choose different names for this new table and the fields in it. 
    If the radial bins are in physical units, the consistency of the input cosmology is checked against any previously defined \code{galcat} cosmology.

    \item \code{plot\_profiles} provides the user with an easy way to plot the tangential and cross profiles from the \code{gc.profile} table and returns a \code{matplotlib.figure} object. 
    This method is provided as a quick diagnostic tool to check of the results of the \code{make\_radial\_profile} operation and is not meant to substitute for user-specific presentation-ready plots.

    \item\code{save} and \code{load} enable the user to keep a \code{GalaxyCluster} object as a \code{pickle} file for later use or retrieve a \code{GalaxyCluster} object from a file.

\end{itemize}

Curation of the membership of a galaxy cluster, for example, based on cuts in redshift or color, as well as measurement uncertainties thereof, is an essential aspect of any analysis. 
Such a procedure may be accomplished by preforming the selection on the inputs to a \code{GalaxyCluster} object, whose \code{galcat} table then corresponds to the subsample.

\subsection{The \code{support} package}
\label{sec:support}

The \clmm\ library also provides the user with a \code{support} package with a few useful functionalities. 

\subsubsection{Mock data generation}
\label{sec:mock_data}

We provide a module to generate a mock galaxy cluster and sample of background galaxies, including the associated shear quantities. 
This has proven useful for testing purposes during the development of the code, but also allows the user to explore some effects that impact the weak lensing cluster mass estimate in a controlled manner. 
The core function of this module is \code{generate\_galaxy\_catalog}, that takes as arguments the redshift, mass and concentration of the cluster to be simulated, specifying also the spherical overdensity at which the mass and concentration are defined, the type of parametric density profile, and the cosmology. 
The cluster is treated as being located at position $(\code{ra},\code{dec}) = (0,0)$. 
Here we describe the options available to the user for the background galaxy sample generation: 
\begin{itemize}
    \item The user sets the size of the field-of-view and the number of background galaxies (or the number density) to be simulated. 
    Each galaxy position is uniformly distributed within the field.
    \item All background galaxies can be set at the same redshift or drawn from the distribution 
    \begin{equation}
        N(z) \propto z^\alpha\exp\left[-\left(\frac{z}{z_0}\right)^\beta\right]
        \label{eq:chang}
    \end{equation}
    over a user-defined redshift range. 
    Pre-defined options available to the user are the \citet{chang2013effective} distribution with the parameters $(\alpha, \beta, z_0) = (1.24, 1.01, 0.51)$ and the \desc\ SRD Y10 \citep{2018arXiv180901669T} distribution with $(\alpha, \beta, z_0) = (2, 0.9, 0.28)$.
    \item Each redshift sampled above corresponds to the true redshift of one mock galaxy. 
    The user can select the option to add redshift errors, and an observed redshift is drawn from a Gaussian distribution with a dispersion of $\sigma_0(1 + z)$, for which the user sets the unscaled standard deviation $\sigma_0$. 
    A probability distribution function for each mock galaxy is also returned. 
    \item Each galaxy is assigned its true shear $g_1+i g_2$ based on its true redshift and position. 
    Shape noise can be added to instead generate the galaxy's lensed ellipticity (see Eq.~\eqref{eq:transform}). 
    The galaxy's intrinsic shape is sampled from a Gaussian distribution centered on 0, with a user-specified dispersion $\sigma_\epsilon$.
    \end{itemize}
The module outputs a \code{GCData} table containing the mock galaxies coordinates, redshifts, lensed ellipticities and shear measurements (plus, optionally, the redshift probability distribution functions). 
The source galaxy mock sample can then be given as an input of the \code{GalaxyCluster} object.

\subsubsection{Optimizers and samplers}
\label{sec:samplers} 

Though the \clmm\ code base does not include optimizer and sampler objects, the \code{support} package does provide straightforward feed-through access to some \code{scipy.optimize} functions in the \code{sampler} module, including \code{curve\_fit}, \code{minimize} and \code{basinhoppin}. 
In some configurations, the latter was found a more robust minimizer than \code{minimize}.

\section{The \clmm\ example suite}
\label{sec:examples} 

The core functions of \clmm\ aim to provide the user with all the tools to self-consistently prepare both the data vector and the models when building an analysis pipeline to fit the cluster mass. 
In the \code{examples} folder, we provide specific notebooks to show all the functionalities of \code{theory}, \code{dataops}, and \code{mock\_data}. 
To maximize future flexibility and minimize maintenance overhead, we choose not to wrap the mass fitting step {\it per se} in the \clmm\ library but instead provide the users with example Jupyter Notebooks highlighting several ways to do so:
\begin{enumerate}
    \item  Example 1 generates mock data from an NFW profile without shape noise or errors in the redshift assuming all galaxies are placed on a single redshift plane and performs a fit on the shear profile, with the $\chi^2$ explicitly written and minimized using \code{scipy.optimize.minimize}.
    \item Example 2 is similar to Example 1 but creates more realistic mock data with galaxies distributed in redshift space, with and without redshift errors and with and without shape noise added, and using \code{scipy.optimize.curve\_fit} to perform the mass fitting. 
    We purposely build the {\it wrong} shear model (ignoring the redshift distribution and using a single source plane at the average redshift) to highlight the resultant bias in the mass reconstruction due to such a misspecification. 
    \item Example 3 uses a similar noisy dataset as in Example 2 and shows one way to account for the galaxy redshift distribution in the shear model to solve the bias highlighted previously in a binned approach. The alternative solution accounts for the redshift distribution in the data vector, working with the $\Delta\Sigma$ profiles instead.
\end{enumerate}
These examples are also designed to assist new scientists entering the field with a pedagogical progression of increasing complexity.

\begin{figure}
  \centering
    \includegraphics[trim=20 20 0 0,clip,width=\columnwidth]{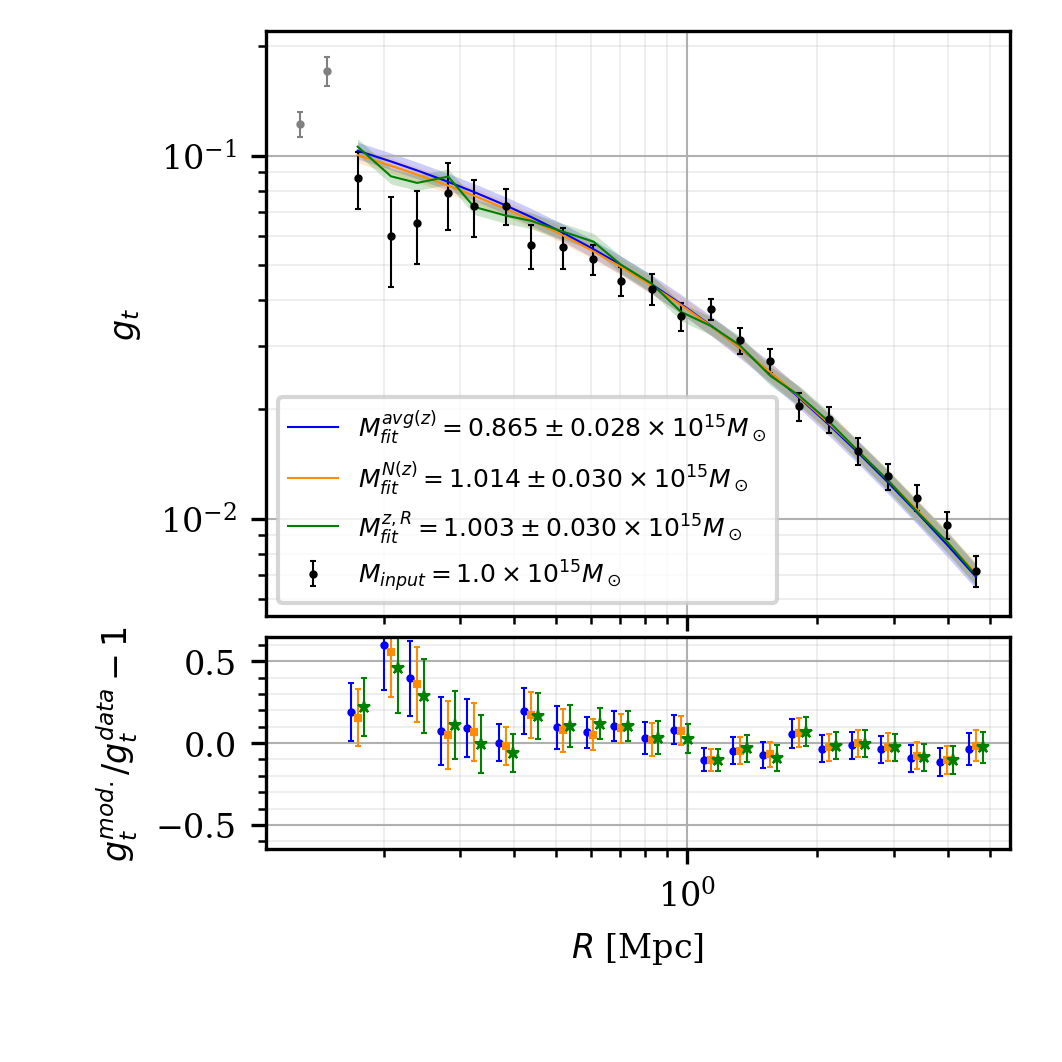}
  \caption{{\it Top:} Tangential reduced shear profiles (symbols) measured from a mock dataset that includes shape noise, redshift distribution, and photometric redshift uncertainties. 
  The solid lines and shaded region correspond to the best-fit model and 3-sigma uncertainty, when fitting for the mass using three different models (assuming a known concentration, see text for details). 
  The gray points correspond to bins with fewer than five galaxies that have not been considered for the fit.
  {\it Bottom:} Residual of each model, with a small shift in radial position for better visualization.
  }
  \label{fig:mock_data} 
\end{figure}

Figure \ref{fig:mock_data} compiles some of these examples together and shows the result of using the mock data module to build a binned shear profile (data points) and the best fit models to these data, using $\chi^2$ minimization.
In this example, the mock data are generated using $\sim 20$ galaxies per square arcmin in a $0.6 \times 0.6$~deg$^2$ field, include realistic photometric redshift errors $\sigma_z = 0.05 \times (1+z)$, and a very low-level of shape noise\footnote{This is an unrealistically low value for the dispersion of intrinsic ellipticities, which is typically $\sim 0.3$. This choice was made so as to allow the narrow bins used in Fig.~\ref{fig:mock_data} which made it unnecessary to integrate the models \eqref{eq:wrong_gt_model} and \eqref{eq:approx_model} over the bin width. The focus of this example is solely that of the inclusion of source redshift distribution in the model.} $\sigma_\epsilon = 0.05$. 
We assume the galaxy shapes in the catalog are ideally measured as the \code{mock\_data} module does not yet support the generation of shape measurement errors.
The fit considers only radial bins of the profile with more than five galaxies, using three different models for the reduced shear described below, with the best-fit mass reported on the figure. 
The first and \emph{naive} theoretical prediction of the reduced 
shear $g_t$ assumes that all sources are at the same redshift, namely the average source redshift. 
The predicted shear in radial bin $i$ (blue line) is given by
 \begin{equation}
     g_{t,i}^{\rm{avg(z)}} = g_t(R_i, \langle z \rangle)\;,
 \label{eq:wrong_gt_model}
 \end{equation} 
where $R_i$ is the average radius of the sources in the bin and $\langle z \rangle$ is the average redshift of all sources.
The second model (orange line) also uses the average radius but accounts for the overall redshift distribution of the sources according to,
 \begin{equation}
     g_{t,i}^{N(z)} = \frac{\langle\beta_s\rangle \gamma_t(R_i, z\rightarrow\infty)}{1-\frac{\langle\beta_s^2\rangle}{\langle\beta_s\rangle}\kappa(R_i, z\rightarrow\infty)} ,
     \label{eq:approx_model}
 \end{equation}
where $\beta_s=\Sigma_{\rm crit}(z_s\rightarrow \infty)/\Sigma_{\rm crit}(z_s)$ and the average is performed over the source redshift distribution, as in, e.g., \citealp{applegate14}. 
Finally, the third model (green line) computes a shear value for each source from their individual redshifts and tangential distance from the cluster center, and then averages these shear values in each radial bin, similarly to how the data vector has been obtained, namely
 \begin{equation}
    g_{t,i}^{z, R} = \frac{1}{N_i}\sum_{{\rm gal\,}j\in {\rm bin\,}i} g_t(R_j, z_j) ,
    \label{eq:exact_model}
 \end{equation}
where $N_i$ is the number of galaxies in bin $i$. 
Because of the different redshift population in each bin, this model is not smooth with radius. 
If no shape noise or redshift errors had been added to the data, the model would be a perfect match to the data.

Looking at the residual (bottom panel), the three models perform similarly. 
However, using the wrong model $g_{t,i}^{\rm{avg(z)}}$ yields a biased mass reconstruction, as seen in the legend. 
The two models accounting for the redshift distribution of the sources (orange and green) yield similar unbiased results.
The notebook used to generate this figure is given in Appendix \ref{sec:api_appendix}, which also serves as a demonstration of some of \clmm's API. 

\begin{figure}
    \centering
    \includegraphics{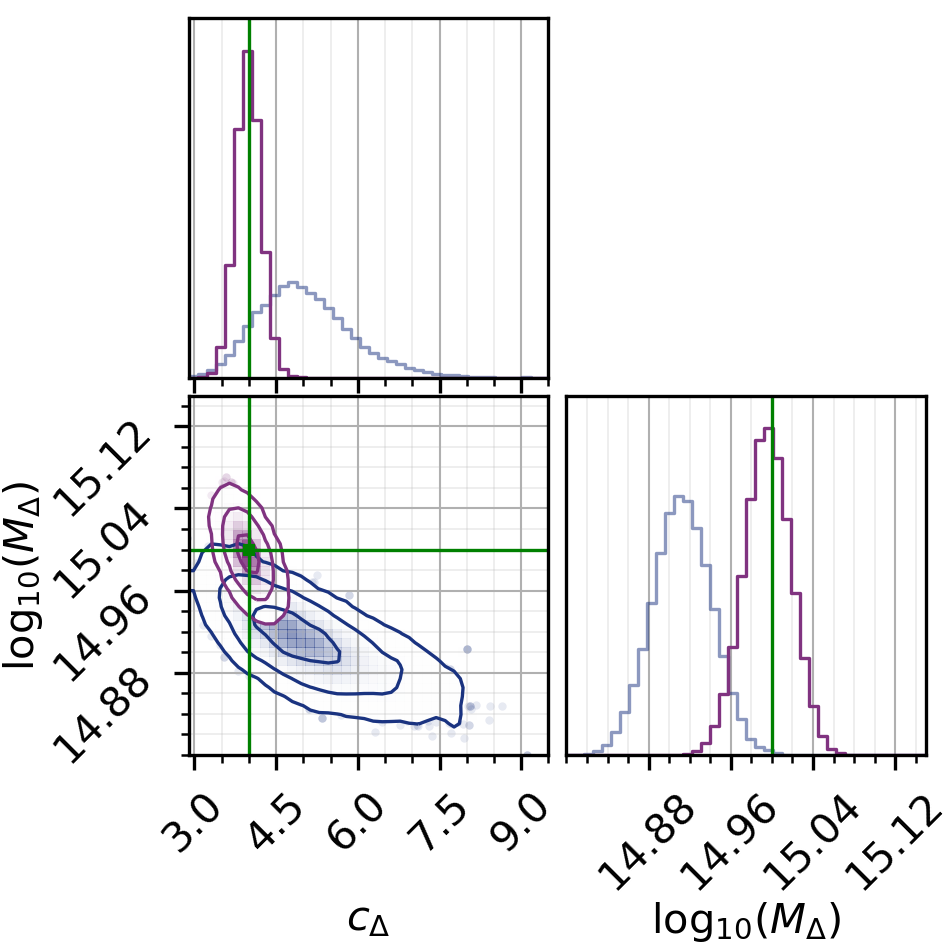}
    \caption{Output the \nc\ version of Example 2, when fitting for both mass and concentration using the MCMC available in the \nc\ statistical framework. In blue, the results are biased when purposely using the wrong model that assumes a single source plane (Eq.\eqref{eq:wrong_gt_model}). The unbiased results of the unbinned likelihood are shown in purple, while the green lines represent the truth.}
    \label{fig:NC_example}
\end{figure}

\paragraph*{NumCosmo-specific examples} 
The above examples will run with any of the modeling \be s described in Table~\ref{tab:backends}. 
However, \nc\ comes with a full statistical environment and Example~1 and Example~2 above are repeated using this framework instead of \code{scipy.optimize}. 
In particular, the \nc\ version of Example 2 shows how to solve the bias introduced by the wrong model of Eq.\eqref{eq:wrong_gt_model} by presenting an unbinned likelihood approach. 
In that approach, the shear measured for each galaxy $\hat{g}_{t,j}$ is used as a data point, allowing the use of the full information available in the data. The final likelihood is the product of the probability distributions of each galaxy $P_j$ given by
\begin{equation}
    P_j = \int_0^\infty
    \frac{1}{\sqrt{2\pi}\sigma}
    \exp\left[- \frac{(\hat{g}_{t,j} - g_t(R_j, z^T))^2)}{2\sigma^2}\right]
    P_j(z^T) \mathrm{d}z^T,
\end{equation} 
where $P_j(z^T)$ is the probability distribution of the true redshift $z^T$ given the photometric data, $R_j$ is assumed to be perfectly known, and $\sigma$ is estimated standard deviation of $\hat{g}_{t,j}$.
It also shows the results when fitting for both mass and concentration using Monte Carlo Markov chains (MCMC), as illustrated in figure~\ref{fig:NC_example}, to highlight, in a different way, the bias introduced when fitting the wrong model (blue) or using the unbinned likelihood approach (purple).

\paragraph*{DC2-specific examples} The \desc\ Data Challenge 2 (DC2) simulated dataset is a major achievement of \desc\ to provide the community with data similar to a \lsst\ multi-year release \citep{korytov2019cosmodc2, abolfathi2020lsst}. 
As part of the \clmm\ examples, we also provide two notebooks showing how to build the shear profiles from the DC2 catalogues. 
These notebooks use access to the catalogs from the \desc\ conda environment at NERSC or CC-IN2P3, but the catalogs have now also been publicly released \citep{2021arXiv210104855L} through the \desc\ data portal\footnote{\url{https://lsstdesc-portal.nersc.gov/}}.

\section{Future directions}
\label{sec:discussion}

In accordance with the scope outlined in Section~\ref{sec:scope}, plans for future versions of \clmm\ are already underway. 
Additional functionalities will be implemented to enable a larger variety of analyses involving the simulation, recovery, and application of galaxy cluster masses.
In this section, we share some specific upcoming functionalities beyond the v1.0 release.
Our plans emphasize the building blocks of flexible analyses accompanied by extensive examples of their use to build a variety of end-to-end pipelines.
Among other things, future examples will encompass fitting individual cluster masses using the algorithm of \citet{applegate14} (anticipated to be accessible through the \nc\ \be\ soon) and that of a stacked shear signal of an ensemble of clusters.  

\subsection{Theory}

Ongoing development of the theory subpackage, both directly in \clmm\ or its \be s, aim to improve the prediction of the weak lensing signal.
We anticipate carrying out informative cross-checks between multiple implementations of these models and sharing those results with the community.

\paragraph*{Halo parameterizations} 
To support a larger variety of halo parametrizations, \clmm\ modeling functionalities could be extended to other dark matter density profiles (e.g. \cite{Burkert_1995}, \citet{Diemer_2014}). This would allow for a more extensive characterization of the halo modeling-related uncertainties when estimating WL cluster masses. For instance, it is well-known that the NFW enclosed mass diverges and, therefore, we need to truncate the halo profile to better describe the transition between the inner and outer regions of the halo (e.g., \citet{Oguri2011}, \citet{Diemer_2014} and references therein).

\paragraph*{Mass-concentration relations}  
Conversely to the example shown in Fig.~\ref{fig:NC_example}, fitting for the WL mass is often performed by assuming a given mass-concentration relation to remove one free parameter from the problem. 
Given a mass definition, numerical simulations have shown that the concentration parameter is scattered around empirical relations linked to the halo mass, cluster redshift, and possibly cosmology (e.g., \citealp{2008MNRAS.390L..64D,2015ApJ...799..108D,2019ApJ...871..168D}). 
Many mass-concentration relations exist in the literature and some are already available in \ccl\ and \nc, which facilitates easy feed-through access from the \be s. 

\paragraph*{Two-halo term} 
On scales larger than $\sim R_{\Delta}$, the lensing signal coming from the contribution of large scale structures (known as the two-halo term) becomes significant and should be accounted for in the modeling to avoid a biased mass reconstruction. 
This contribution depends on the matter power spectrum and is implemented in \ct. 
Elements to compute this term already exist in \nc\ and \ccl, but the full implementation is yet to be done in those \be s. 
Once this functionality is available and can be cross-checked for at least two of the \clmm\ \be s, we will make it available in \clmm. 

\paragraph*{Miscentering} 
Another issue when modeling the mass distribution of clusters is the choice of the center of the mass distribution provided to the model, which is usually estimated from baryonic tracers, such as the central galaxy or the peak of the X-ray luminosity. 
These observations may not match the center of the mass distribution with sufficient precision or accuracy\footnote{Note also that the use of a halo center to make a radial profile is only valid when assuming spherical symmetry, subdominant substructure, and a negligible merger rate, which does not occur in reality. A future version of \clmm may include more sophisticated functionality beyond the traditional azimuthal averaging approach to handle these more complex situations.}, often leading to an underestimated mass \citep[e.g.][]{hoekstra11} as the lensing signal in the inner regions is suppressed. 
Procedures to correct the shear model for miscentering (e.g., \citealp{2019MNRAS.482.1352M}) will be included in the next version of \clmm.

\subsection{Data operations}

On the data analysis side, the current release focuses on building the shear profile of individual clusters. 
This will soon be supplemented by the following functionalities.

\paragraph*{Cluster ensembles} 
As individual galaxy cluster shear signals may be noise-dominated, particularly in the faint regime probed by \lsst, cosmological constraints in cluster lensing are typically derived from large samples of galaxy clusters rather than only the subpopulation of massive clusters for which a precise mass measurement is accessible.
By combining data from many clusters subject to the same cosmology and accounting for diversity in redshift and richness, the mean cluster properties may be estimated with high signal-to-noise ratio, facilitating competitive cosmological constraints utilizing clusters at a broader range of mass.
Built upon the \code{GalaxyCluster} object, work has begun on a \code{ClusterEnsemble} object representing a collection of multiple galaxy clusters.
Such an object enables complete cosmological analyses involving samples of galaxy clusters, from widely-used stacked shear signal estimation approaches (e.g, \citealp{2019MNRAS.482.1352M}) to more sophisticated hierarchical Bayesian inference procedures \citep{2017MNRAS.468.4872L}, demonstrations of which will accompany the introduction of the \code{ClusterEnsemble} object.

\paragraph*{Magnification signal}
So far \clmm\ focuses only on shear as an observable. 
However, the gravitational magnification signal in the weak lensing regime is a promising alternate probe of clusters masses \citep[see][for a recent example]{2020MNRAS.495..428C}. 
Furthermore, the magnification and shear signal can be predicted and fitted jointly, to increase the signal-to-noise ratio \citep{2014ApJ...795..163U}. 
While the magnification predictions are already available in the \code{theory} module of \clmm, upcoming versions of \clmm\ will include weak lensing magnification observables (namely the magnification bias, i.e. the density contrast of background sources behind lenses) in the \code{dataops} package and the \code{mock\_data} generation module.

\subsection{Handling of systematic uncertainties}
Several sources of systematic uncertainty affect cluster weak lensing mass estimates, coming from both data and modeling assumptions (see e.g., \citealp{2017MNRAS.469.4899M}). Careful accounting of how these propagates from the shear profile to the mass is fundamental for cluster-based cosmology. While some effects may be parameterized and included in the fit as nuisance parameters (e.g. miscentering, dilution of the model to account for cluster member contamination), others need to be calibrated on simulations (e.g., modeling systematics coming from the choice of profile, mass-concentration relation or the assumption of spherical symmetry). Future developments of \clmm will include tools for parametrization and calibration that facilitate these analyses.


\subsection{Interfacing with other \desc\ tools}
In addition to being a standalone library open to the community to compute WL cluster masses, \clmm\ is part of the \desc\ effort to produce robust software infrastructure enabling flexible multi-probe cosmology analyses. 
The latter consideration motivated the inclusion of the \desc\ Core Cosmology Library (\ccl) as one of \clmm's \be s. 
Interfacing with other \desc\ tools, such as 
\code{FireCrown}\footnote{\url{https://github.com/LSSTDESC/firecrown}}, 
\code{qp}\footnote{\url{https://github.com/LSSTDESC/qp}}, 
or \code{TXPipe}\footnote{\url{https://github.com/LSSTDESC/TXPipe}} 
will become even more important in the near future as the development of a \desc\ cluster cosmology pipeline built on \clmm progresses.

\section{Summary}
\label{sec:summary}

In this paper, we present the first release of \clmm, an open-source library developed to provide users with the necessary ingredients with which to build a software pipeline performing galaxy cluster weak lensing mass estimation. 
It has been developed as a toolkit for the forthcoming \desc\ cluster cosmology analysis but also serves as a standalone package for the broader cluster lensing community. 

The \clmm\ library relies on existing public codes as \be s for the theoretical predictions of cluster weak lensing quantities. 
As theory \be s, \clmm\ supports \code{cluster-toolkit}, \code{CCL} and \code{NumCosmo}, all of which have been carefully cross-checked, each with their own modeling options for density profiles, mass definition, and cosmology objects users interact with through a \code{Cosmology} wrapper within \clmm.

Along with the theoretical predictions, \clmm\ also provides functionality for performing data operations aimed at preparing the data vector for cluster weak lensing analyses. 
For this first version of the code, these focus on building the gravitational lensing shear or excess surface density profiles of individual clusters. 
These operations are methods of the \code{GalaxyCluster} object that contains all the information related to a given cluster, from the catalog of background galaxies, to the radial lensing profiles. 
\clmm's \code{GalaxyCluster} object ensures consistency of assumptions, such as that of a model cosmology, through sequential stages of an analysis pipeline built with \clmm.

\clmm\ is open-source and has been developed using best coding practices including code reviews, unit tests, validation tests, and continuous integration, to provide users with a robust, fully-tested, and fully-documented code base and development community that actively welcomes new contributors. 
\clmm\ also contains an extensive example suite with progressing complexity aimed at displaying use cases as well as helping scientists new to the field to get started in a pedagogical manner. 
These are generally based on mock data sets, generated by a dedicated \clmm\ \code{support} module, that allow us to mimic some of the uncertainties of cluster weak lensing mass measurements. 

\clmm\ is currently being used in several analyses, e.g., using the \desc\ DC2 simulated data (an example of which is available in the DESC cosmoDC2 validation paper - ref to be added once available) and \collab{DECam} cluster observations reprocessed by the \collab{Rubin} science pipeline. 
\clmm\ Version 1.0 has been designed as a robust basis to encourage and facilitate the implementation of new functionalities, which are being developed continuously and publicly, that will include a much wider range of options on both the theory and data operations sides.


\subsection*{Data availability}
Data used in this paper is generated by the {\tt mock\_data} module of the \clmm code, publicly available online at \url{https://github.com/LSSTDESC/clmm}.

\subsection*{Acknowledgments}

Author contributions are listed below. \\
M.~Aguena: conceptualization, methodology, software, validation, writing -- review \& editing \\
C.~Avestruz: conceptualization, funding acquisition, methodology, project administration, software, validation, writing -- original draft, writing -- review \& editing \\
C.~Combet: conceptualization, methodology, project administration, software, validation, writing -- original draft, writing -- review \& editing \\
S.~Fu: software, writing -- original draft, writing -- review \& editing \\
R.~Herbonnet: software, writing -- original draft, writing -- review \& editing \\
A.I.~Malz: conceptualization, funding acquisition, methodology, project administration, software, validation, writing -- original draft, writing -- review \& editing \\
M.~Penna-Lima: software, validation, writing -- original draft, writing -- review \& editing \\
M.~Ricci: conceptualization, software, methodology, validation, writing -- original draft, writing -- review \& editing \\
S.~D.~P.~Vitenti: conceptualization, methodology, software, validation \\
L.~Baumont: software \\
H.~Fan: software \\
M.~Fong: software \\
M.~Ho: software, writing - original draft \\
M.~Kirby: methodology, software, validation \\
C.~Payerne: software, writing - original draft, writing -- review \& editing \\
D.~Boutigny: conceptualization, investigation \\
B.~Lee: software \\
B.~Liu: software, validation \\
T.~McClintock: software, validation \\
H.~Miyatake: conceptualization \\
C.~Sif\'on: conceptualization, methodology, software \\
A.~von der Linden: conceptualization, methodology, supervision \\
H.~Wu: conceptualization, methodology, software \\
M.~Yoon: software \\

This paper has undergone internal review in the LSST Dark Energy Science Collaboration. The authors would like to thank Elisa Chisari, Douglas Clowe and Ian Dell'Antonio for serving as the LSST-DESC publication review committee whose comments and suggestions that improved the quality of this manuscript. The authors further wish to thank Mike Jarvis for feedback on the early architecture and development decisions of CLMM and David Alonso for his help with testing \ccl's integration of the profiles.

The DESC acknowledges ongoing support from the Institut National de Physique Nucl\'eaire et de Physique des Particules in France; the Science \& Technology Facilities Council in the United Kingdom; and the Department of Energy, the National Science Foundation, and the LSST Corporation in the United States.  DESC uses resources of the IN2P3 Computing Center (CC-IN2P3--Lyon/Villeurbanne - France) funded by the Centre National de la Recherche Scientifique; the National Energy Research Scientific Computing Center, a DOE Office of Science User Facility supported by the Office of Science of the U.S.\ Department of Energy under Contract No.\ DE-AC02-05CH11231; STFC DiRAC HPC Facilities, funded by UK BIS National E-infrastructure capital grants; and the UK particle physics grid, supported by the GridPP Collaboration.  This work was performed in part under DOE Contract DE-AC02-76SF00515.

The authors express gratitude to the LSSTC for the 2018 and 2019 Enabling Science grants, hosted by CMU and RUB respectively, that supported the development of \clmm and its developer community.

CA acknowledges support from the LSA Collegiate Fellowship at the University of Michigan, the Leinweber Foundation, and DoE Award DE-FOA-0001781.

AIM acknowledges support from the Max Planck Society and the Alexander von Humboldt Foundation in the framework of the Max Planck-Humboldt Research Award endowed by the Federal Ministry of Education and Research. During the completion of this work, AIM was advised by David W. Hogg and supported by National Science Foundation grant AST-1517237.
CS acknowledges support from the Agencia Nacional de Investigaci\'on y Desarrollo (ANID) through FONDECYT grant no.\ 11191125.

AvdL, RH, LB, and HF acknowledge support by the US Department of Energy under award DE-SC0018053.

SF acknowledges support from DOE grant DE-SC0010010.

HM is supported by the Jet Propulsion Laboratory, California Institute of Technology, under a contract with the National Aeronautics and Space Administration.

\bibliographystyle{mnras}
\bibliography{main}

\begin{thebibliography}{}
\makeatletter
\relax
\def\mn@urlcharsother{\let\do\@makeother \do\$\do\&\do\#\do\^\do\_\do\%\do\~}
\def\mn@doi{\begingroup\mn@urlcharsother \@ifnextchar [ {\mn@doi@}
  {\mn@doi@[]}}
\def\mn@doi@[#1]#2{\def\@tempa{#1}\ifx\@tempa\@empty \href
  {http://dx.doi.org/#2} {doi:#2}\else \href {http://dx.doi.org/#2} {#1}\fi
  \endgroup}
\def\mn@eprint#1#2{\mn@eprint@#1:#2::\@nil}
\def\mn@eprint@arXiv#1{\href {http://arxiv.org/abs/#1} {{\tt arXiv:#1}}}
\def\mn@eprint@dblp#1{\href {http://dblp.uni-trier.de/rec/bibtex/#1.xml}
  {dblp:#1}}
\def\mn@eprint@#1:#2:#3:#4\@nil{\def\@tempa {#1}\def\@tempb {#2}\def\@tempc
  {#3}\ifx \@tempc \@empty \let \@tempc \@tempb \let \@tempb \@tempa \fi \ifx
  \@tempb \@empty \def\@tempb {arXiv}\fi \@ifundefined
  {mn@eprint@\@tempb}{\@tempb:\@tempc}{\expandafter \expandafter \csname
  mn@eprint@\@tempb\endcsname \expandafter{\@tempc}}}

\bibitem[\protect\citeauthoryear{{Abbott} et~al.,}{{Abbott}
  et~al.}{2020}]{2020PhRvD.102b3509A}
{Abbott} T.~M.~C.,  et~al., 2020, \mn@doi [\prd] {10.1103/PhysRevD.102.023509},
  \href {https://ui.adsabs.harvard.edu/abs/2020PhRvD.102b3509A} {102, 023509}

\bibitem[\protect\citeauthoryear{Abolfathi et~al.,}{Abolfathi
  et~al.}{2020}]{abolfathi2020lsst}
Abolfathi B.,  et~al., 2020, arXiv preprint arXiv:2010.05926

\bibitem[\protect\citeauthoryear{{Ansarifard} et~al.,}{{Ansarifard}
  et~al.}{2020}]{ansarifardetal20}
{Ansarifard} S.,  et~al., 2020, \mn@doi [\aap] {10.1051/0004-6361/201936742},
  \href {https://ui.adsabs.harvard.edu/abs/2020A&A...634A.113A} {634, A113}

\bibitem[\protect\citeauthoryear{{Applegate} et~al.,}{{Applegate}
  et~al.}{2014}]{applegate14}
{Applegate} D.~E.,  et~al., 2014, \mn@doi [\mnras] {10.1093/mnras/stt2129},
  \href {https://ui.adsabs.harvard.edu/abs/2014MNRAS.439...48A} {439, 48}

\bibitem[\protect\citeauthoryear{{Bartelmann} \& {Schneider}}{{Bartelmann} \&
  {Schneider}}{2001}]{bartelmannandschneider2001}
{Bartelmann} M.,  {Schneider} P.,  2001, \mn@doi [\physrep]
  {10.1016/S0370-1573(00)00082-X}, \href
  {https://ui.adsabs.harvard.edu/abs/2001PhR...340..291B} {340, 291}

\bibitem[\protect\citeauthoryear{{Becker} \& {Kravtsov}}{{Becker} \&
  {Kravtsov}}{2011}]{beckerandkravtsov11}
{Becker} M.~R.,  {Kravtsov} A.~V.,  2011, \mn@doi [\apj]
  {10.1088/0004-637X/740/1/25}, \href
  {https://ui.adsabs.harvard.edu/abs/2011ApJ...740...25B} {740, 25}

\bibitem[\protect\citeauthoryear{{Bryan} \& {Norman}}{{Bryan} \&
  {Norman}}{1998}]{Bryan1998}
{Bryan} G.~L.,  {Norman} M.~L.,  1998, \mn@doi [\apj] {10.1086/305262}, \href
  {https://ui.adsabs.harvard.edu/abs/1998ApJ...495...80B} {495, 80}

\bibitem[\protect\citeauthoryear{Burkert}{Burkert}{1995}]{Burkert_1995}
Burkert A.,  1995, \mn@doi [The Astrophysical Journal] {10.1086/309560}, 447

\bibitem[\protect\citeauthoryear{Chang et~al.,}{Chang
  et~al.}{2013}]{chang2013effective}
Chang C.,  et~al., 2013, Monthly Notices of the Royal Astronomical Society,
  434, 2121

\bibitem[\protect\citeauthoryear{{Chisari} et~al.,}{{Chisari}
  et~al.}{2019}]{2019ApJS..242....2C}
{Chisari} N.~E.,  et~al., 2019, \mn@doi [\apjs] {10.3847/1538-4365/ab1658},
  \href {https://ui.adsabs.harvard.edu/abs/2019ApJS..242....2C} {242, 2}

\bibitem[\protect\citeauthoryear{{Chiu}, {Umetsu}, {Murata}, {Medezinski}  \&
  {Oguri}}{{Chiu} et~al.}{2020}]{2020MNRAS.495..428C}
{Chiu} I.~N.,  {Umetsu} K.,  {Murata} R.,  {Medezinski} E.,   {Oguri} M.,
  2020, \mn@doi [\mnras] {10.1093/mnras/staa1158}, \href
  {https://ui.adsabs.harvard.edu/abs/2020MNRAS.495..428C} {495, 428}

\bibitem[\protect\citeauthoryear{{Corless} \& {King}}{{Corless} \&
  {King}}{2007}]{corless07}
{Corless} V.~L.,  {King} L.~J.,  2007, \mn@doi [\mnras]
  {10.1111/j.1365-2966.2007.12018.x}, \href
  {https://ui.adsabs.harvard.edu/abs/2007MNRAS.380..149C} {380, 149}

\bibitem[\protect\citeauthoryear{{Dias Pinto Vitenti} \& {Penna-Lima}}{{Dias
  Pinto Vitenti} \& {Penna-Lima}}{2014}]{2014ascl.soft08013D}
{Dias Pinto Vitenti} S.,  {Penna-Lima} M.,  2014, {NumCosmo: Numerical
  Cosmology} (\mn@eprint {ascl} {1408.013})

\bibitem[\protect\citeauthoryear{{Diemer} \& {Joyce}}{{Diemer} \&
  {Joyce}}{2019}]{2019ApJ...871..168D}
{Diemer} B.,  {Joyce} M.,  2019, \mn@doi [\apj] {10.3847/1538-4357/aafad6},
  \href {https://ui.adsabs.harvard.edu/abs/2019ApJ...871..168D} {871, 168}

\bibitem[\protect\citeauthoryear{Diemer \& Kravtsov}{Diemer \&
  Kravtsov}{2014}]{Diemer_2014}
Diemer B.,  Kravtsov A.~V.,  2014, \mn@doi [The Astrophysical Journal]
  {10.1088/0004-637x/789/1/1}, 789, 1

\bibitem[\protect\citeauthoryear{{Diemer} \& {Kravtsov}}{{Diemer} \&
  {Kravtsov}}{2015}]{2015ApJ...799..108D}
{Diemer} B.,  {Kravtsov} A.~V.,  2015, \mn@doi [\apj]
  {10.1088/0004-637X/799/1/108}, \href
  {https://ui.adsabs.harvard.edu/abs/2015ApJ...799..108D} {799, 108}

\bibitem[\protect\citeauthoryear{{Dietrich} et~al.,}{{Dietrich}
  et~al.}{2014}]{dietrich14}
{Dietrich} J.~P.,  et~al., 2014, \mn@doi [\mnras] {10.1093/mnras/stu1282},
  \href {https://ui.adsabs.harvard.edu/abs/2014MNRAS.443.1713D} {443, 1713}

\bibitem[\protect\citeauthoryear{{Dodelson}, {Heitmann}, {Hirata}, {Honscheid},
  {Roodman}, {Seljak}, {Slosar}  \& {Trodden}}{{Dodelson}
  et~al.}{2016}]{dodelson16}
{Dodelson} S.,  {Heitmann} K.,  {Hirata} C.,  {Honscheid} K.,  {Roodman} A.,
  {Seljak} U.,  {Slosar} A.,   {Trodden} M.,  2016, arXiv e-prints, \href
  {https://ui.adsabs.harvard.edu/abs/2016arXiv160407626D} {p. arXiv:1604.07626}

\bibitem[\protect\citeauthoryear{{Duffy}, {Schaye}, {Kay}  \& {Dalla
  Vecchia}}{{Duffy} et~al.}{2008}]{2008MNRAS.390L..64D}
{Duffy} A.~R.,  {Schaye} J.,  {Kay} S.~T.,   {Dalla Vecchia} C.,  2008, \mn@doi
  [\mnras] {10.1111/j.1745-3933.2008.00537.x}, \href
  {https://ui.adsabs.harvard.edu/abs/2008MNRAS.390L..64D} {390, L64}

\bibitem[\protect\citeauthoryear{Einasto}{Einasto}{1965}]{einasto1965}
Einasto J.,  1965, Trudy Astrofizicheskogo Instituta Alma-Ata, \href
  {https://ui.adsabs.harvard.edu/abs/1965TrAlm...5...87E} {5, 87}

\bibitem[\protect\citeauthoryear{{Ford} \& {VanderPlas}}{{Ford} \&
  {VanderPlas}}{2016}]{2016AJ....152..228F}
{Ford} J.,  {VanderPlas} J.,  2016, \mn@doi [\aj]
  {10.3847/1538-3881/152/6/228}, \href
  {https://ui.adsabs.harvard.edu/abs/2016AJ....152..228F} {152, 228}

\bibitem[\protect\citeauthoryear{Harris et~al.,}{Harris
  et~al.}{2020}]{harris_numpy_2020}
Harris C.~R.,  et~al., 2020, \mn@doi [Nature] {10.1038/s41586-020-2649-2}, 585,
  357–362

\bibitem[\protect\citeauthoryear{{Henson}, {Barnes}, {Kay}, {McCarthy}  \&
  {Schaye}}{{Henson} et~al.}{2017}]{hensonetal17}
{Henson} M.~A.,  {Barnes} D.~J.,  {Kay} S.~T.,  {McCarthy} I.~G.,   {Schaye}
  J.,  2017, \mn@doi [\mnras] {10.1093/mnras/stw2899}, \href
  {https://ui.adsabs.harvard.edu/abs/2017MNRAS.465.3361H} {465, 3361}

\bibitem[\protect\citeauthoryear{{Herbonnet} et~al.,}{{Herbonnet}
  et~al.}{2020}]{herbonnetetal20}
{Herbonnet} R.,  et~al., 2020, \mn@doi [\mnras] {10.1093/mnras/staa2303}, \href
  {https://ui.adsabs.harvard.edu/abs/2020MNRAS.497.4684H} {497, 4684}

\bibitem[\protect\citeauthoryear{{Hernandez-Martin} et~al.,}{{Hernandez-Martin}
  et~al.}{2020}]{hernandez20}
{Hernandez-Martin} B.,  et~al., 2020, arXiv e-prints, \href
  {https://ui.adsabs.harvard.edu/abs/2020arXiv200700386H} {p. arXiv:2007.00386}

\bibitem[\protect\citeauthoryear{Hernquist}{Hernquist}{1990}]{hernquist1990}
Hernquist L.,  1990, \mn@doi [The Astrophysical Journal] {10.1086/168845},
  \href {https://ui.adsabs.harvard.edu/abs/1990ApJ...356..359H} {356, 359}

\bibitem[\protect\citeauthoryear{{Hoekstra}}{{Hoekstra}}{2003}]{hoekstra03}
{Hoekstra} H.,  2003, \mn@doi [\mnras] {10.1046/j.1365-8711.2003.06264.x},
  \href {https://ui.adsabs.harvard.edu/abs/2003MNRAS.339.1155H} {339, 1155}

\bibitem[\protect\citeauthoryear{{Hoekstra}, {Donahue}, {Conselice}, {McNamara}
   \& {Voit}}{{Hoekstra} et~al.}{2011}]{hoekstra11}
{Hoekstra} H.,  {Donahue} M.,  {Conselice} C.~J.,  {McNamara} B.~R.,   {Voit}
  G.~M.,  2011, \mn@doi [\apj] {10.1088/0004-637X/726/1/48}, \href
  {https://ui.adsabs.harvard.edu/abs/2011ApJ...726...48H} {726, 48}

\bibitem[\protect\citeauthoryear{Hunter}{Hunter}{2007}]{hunter_matplotlib:_2007}
Hunter J.~D.,  2007, \mn@doi [Computing in Science Engineering]
  {10.1109/MCSE.2007.55}, 9, 90

\bibitem[\protect\citeauthoryear{{Kannawadi} et~al.,}{{Kannawadi}
  et~al.}{2019}]{kannawadi19}
{Kannawadi} A.,  et~al., 2019, \mn@doi [\aap] {10.1051/0004-6361/201834819},
  \href {https://ui.adsabs.harvard.edu/abs/2019A&A...624A..92K} {624, A92}

\bibitem[\protect\citeauthoryear{Korytov et~al.,}{Korytov
  et~al.}{2019}]{korytov2019cosmodc2}
Korytov D.,  et~al., 2019, The Astrophysical Journal Supplement Series, 245, 26

\bibitem[\protect\citeauthoryear{{LSST Dark Energy Science
  Collaboration}}{{LSST Dark Energy Science Collaboration}}{2012}]{lsst2012}
{LSST Dark Energy Science Collaboration} 2012, arXiv e-prints, \href
  {https://ui.adsabs.harvard.edu/abs/2012arXiv1211.0310L} {p. arXiv:1211.0310}

\bibitem[\protect\citeauthoryear{{LSST Dark Energy Science Collaboration}
  et~al.,}{{LSST Dark Energy Science Collaboration}
  et~al.}{2021}]{2021arXiv210104855L}
{LSST Dark Energy Science Collaboration} et~al., 2021, arXiv e-prints, \href
  {https://ui.adsabs.harvard.edu/abs/2021arXiv210104855L} {p. arXiv:2101.04855}

\bibitem[\protect\citeauthoryear{{Lesci} et~al.,}{{Lesci}
  et~al.}{2020}]{2020arXiv201212273L}
{Lesci} G.~F.,  et~al., 2020, arXiv e-prints, \href
  {https://ui.adsabs.harvard.edu/abs/2020arXiv201212273L} {p. arXiv:2012.12273}

\bibitem[\protect\citeauthoryear{{Lieu}, {Farr}, {Betancourt}, {Smith},
  {Sereno}  \& {McCarthy}}{{Lieu} et~al.}{2017}]{2017MNRAS.468.4872L}
{Lieu} M.,  {Farr} W.~M.,  {Betancourt} M.,  {Smith} G.~P.,  {Sereno} M.,
  {McCarthy} I.~G.,  2017, \mn@doi [\mnras] {10.1093/mnras/stx686}, \href
  {https://ui.adsabs.harvard.edu/abs/2017MNRAS.468.4872L} {468, 4872}

\bibitem[\protect\citeauthoryear{{Mamajek} et~al.,}{{Mamajek}
  et~al.}{2015}]{2015arXiv151007674M}
{Mamajek} E.~E.,  et~al., 2015, arXiv e-prints, \href
  {https://ui.adsabs.harvard.edu/abs/2015arXiv151007674M} {p. arXiv:1510.07674}

\bibitem[\protect\citeauthoryear{{Mandelbaum} et~al.,}{{Mandelbaum}
  et~al.}{2015}]{great3}
{Mandelbaum} R.,  et~al., 2015, \mn@doi [\mnras] {10.1093/mnras/stv781}, \href
  {https://ui.adsabs.harvard.edu/abs/2015MNRAS.450.2963M} {450, 2963}

\bibitem[\protect\citeauthoryear{{Mandelbaum} et~al.,}{{Mandelbaum}
  et~al.}{2018}]{mandelbaum18}
{Mandelbaum} R.,  et~al., 2018, \mn@doi [\mnras] {10.1093/mnras/sty2420}, \href
  {https://ui.adsabs.harvard.edu/abs/2018MNRAS.481.3170M} {481, 3170}

\bibitem[\protect\citeauthoryear{{McClintock} et~al.,}{{McClintock}
  et~al.}{2019}]{2019MNRAS.482.1352M}
{McClintock} T.,  et~al., 2019, \mn@doi [\mnras] {10.1093/mnras/sty2711}, \href
  {https://ui.adsabs.harvard.edu/abs/2019MNRAS.482.1352M} {482, 1352}

\bibitem[\protect\citeauthoryear{{Medezinski} et~al.,}{{Medezinski}
  et~al.}{2018}]{medezinski18}
{Medezinski} E.,  et~al., 2018, \mn@doi [\pasj] {10.1093/pasj/psy009}, \href
  {https://ui.adsabs.harvard.edu/abs/2018PASJ...70...30M} {70, 30}

\bibitem[\protect\citeauthoryear{{Melchior} et~al.,}{{Melchior}
  et~al.}{2017}]{2017MNRAS.469.4899M}
{Melchior} P.,  et~al., 2017, \mn@doi [\mnras] {10.1093/mnras/stx1053}, \href
  {https://ui.adsabs.harvard.edu/abs/2017MNRAS.469.4899M} {469, 4899}

\bibitem[\protect\citeauthoryear{{Meneghetti}, {Rasia}, {Merten}, {Bellagamba},
  {Ettori}, {Mazzotta}, {Dolag}  \& {Marri}}{{Meneghetti}
  et~al.}{2010}]{meneghetti10}
{Meneghetti} M.,  {Rasia} E.,  {Merten} J.,  {Bellagamba} F.,  {Ettori} S.,
  {Mazzotta} P.,  {Dolag} K.,   {Marri} S.,  2010, \mn@doi [\aap]
  {10.1051/0004-6361/200913222}, \href
  {https://ui.adsabs.harvard.edu/abs/2010A&A...514A..93M} {514, A93}

\bibitem[\protect\citeauthoryear{Navarro, Frenk  \& White}{Navarro
  et~al.}{1997}]{NFW1997}
Navarro J.~F.,  Frenk C.~S.,   White S. D.~M.,  1997, \mn@doi [The
  Astrophysical Journal] {10.1086/304888}, \href
  {https://ui.adsabs.harvard.edu/abs/1997ApJ...490..493N} {490, 493}

\bibitem[\protect\citeauthoryear{{Nelson}, {Lau}, {Nagai}, {Rudd}  \&
  {Yu}}{{Nelson} et~al.}{2014}]{nelsonetal14}
{Nelson} K.,  {Lau} E.~T.,  {Nagai} D.,  {Rudd} D.~H.,   {Yu} L.,  2014,
  \mn@doi [\apj] {10.1088/0004-637X/782/2/107}, \href
  {https://ui.adsabs.harvard.edu/abs/2014ApJ...782..107N} {782, 107}

\bibitem[\protect\citeauthoryear{{Oguri} \& {Hamana}}{{Oguri} \&
  {Hamana}}{2011}]{Oguri2011}
{Oguri} M.,  {Hamana} T.,  2011, \mn@doi [\mnras]
  {10.1111/j.1365-2966.2011.18481.x}, \href
  {https://ui.adsabs.harvard.edu/abs/2011MNRAS.414.1851O} {414, 1851}

\bibitem[\protect\citeauthoryear{{Piffaretti} \& {Valdarnini}}{{Piffaretti} \&
  {Valdarnini}}{2008}]{piffarettiandvaldarnini08}
{Piffaretti} R.,  {Valdarnini} R.,  2008, \mn@doi [\aap]
  {10.1051/0004-6361:200809739}, \href
  {https://ui.adsabs.harvard.edu/abs/2008A&A...491...71P} {491, 71}

\bibitem[\protect\citeauthoryear{{Planck Collaboration} et~al.,}{{Planck
  Collaboration} et~al.}{2020}]{planckcosmology18}
{Planck Collaboration} et~al., 2020, \mn@doi [\aap]
  {10.1051/0004-6361/201833910}, \href
  {https://ui.adsabs.harvard.edu/abs/2020A&A...641A...6P} {641, A6}

\bibitem[\protect\citeauthoryear{{Price-Whelan} et~al.,}{{Price-Whelan}
  et~al.}{2018}]{astropy:2018}
{Price-Whelan} A.~M.,  et~al., 2018, \mn@doi [\aj] {10.3847/1538-3881/aabc4f},
  \href {https://ui.adsabs.harvard.edu/#abs/2018AJ....156..123T} {156, 123}

\bibitem[\protect\citeauthoryear{{Sheldon} \& {Huff}}{{Sheldon} \&
  {Huff}}{2017}]{metacal}
{Sheldon} E.~S.,  {Huff} E.~M.,  2017, \mn@doi [\apj]
  {10.3847/1538-4357/aa704b}, \href
  {https://ui.adsabs.harvard.edu/abs/2017ApJ...841...24S} {841, 24}

\bibitem[\protect\citeauthoryear{{Sheldon} et~al.,}{{Sheldon}
  et~al.}{2004}]{2004AJ....127.2544S}
{Sheldon} E.~S.,  et~al., 2004, \mn@doi [\aj] {10.1086/383293}, \href
  {https://ui.adsabs.harvard.edu/abs/2004AJ....127.2544S} {127, 2544}

\bibitem[\protect\citeauthoryear{{Sunayama} et~al.,}{{Sunayama}
  et~al.}{2020}]{sunayama20}
{Sunayama} T.,  et~al., 2020, \mn@doi [\mnras] {10.1093/mnras/staa1646}, \href
  {https://ui.adsabs.harvard.edu/abs/2020MNRAS.496.4468S} {496, 4468}

\bibitem[\protect\citeauthoryear{{The LSST Dark Energy Science Collaboration}
  et~al.,}{{The LSST Dark Energy Science Collaboration}
  et~al.}{2018}]{2018arXiv180901669T}
{The LSST Dark Energy Science Collaboration} et~al., 2018, arXiv e-prints,
  \href {https://ui.adsabs.harvard.edu/abs/2018arXiv180901669T} {p.
  arXiv:1809.01669}

\bibitem[\protect\citeauthoryear{{Tiesinga}, {Mohr}, {Newell}  \&
  {Taylor}}{{Tiesinga} et~al.}{2020}]{codata18}
{Tiesinga} E.,  {Mohr} P.~J.,  {Newell} D.~B.,   {Taylor} B.~N.,  2020, The
  2018 CODATA Recommended Values of the Fundamental Physical Constants, \url
  {https://physics.nist.gov/cuu/Constants/index.html}

\bibitem[\protect\citeauthoryear{{To} et~al.,}{{To}
  et~al.}{2020}]{2020arXiv201001138T}
{To} C.,  et~al., 2020, arXiv e-prints, \href
  {https://ui.adsabs.harvard.edu/abs/2020arXiv201001138T} {p. arXiv:2010.01138}

\bibitem[\protect\citeauthoryear{{Umetsu}}{{Umetsu}}{2020}]{2020A&ARv..28....7U}
{Umetsu} K.,  2020, \mn@doi [\aapr] {10.1007/s00159-020-00129-w}, \href
  {https://ui.adsabs.harvard.edu/abs/2020A&ARv..28....7U} {28, 7}

\bibitem[\protect\citeauthoryear{{Umetsu} et~al.,}{{Umetsu}
  et~al.}{2014}]{2014ApJ...795..163U}
{Umetsu} K.,  et~al., 2014, \mn@doi [\apj] {10.1088/0004-637X/795/2/163}, \href
  {https://ui.adsabs.harvard.edu/abs/2014ApJ...795..163U} {795, 163}

\bibitem[\protect\citeauthoryear{{Varga} et~al.,}{{Varga}
  et~al.}{2019}]{varga19}
{Varga} T.~N.,  et~al., 2019, \mn@doi [\mnras] {10.1093/mnras/stz2185}, \href
  {https://ui.adsabs.harvard.edu/abs/2019MNRAS.489.2511V} {489, 2511}

\bibitem[\protect\citeauthoryear{{Virtanen} et~al.,}{{Virtanen}
  et~al.}{2020}]{virtanen_scipy_2020}
{Virtanen} P.,  et~al., 2020, \mn@doi [Nature Methods]
  {10.1038/s41592-019-0686-2}, \href
  {https://ui.adsabs.harvard.edu/abs/2020NatMe..17..261V} {17, 261}

\bibitem[\protect\citeauthoryear{{Wright} \& {Brainerd}}{{Wright} \&
  {Brainerd}}{2000}]{wright00}
{Wright} C.~O.,  {Brainerd} T.~G.,  2000, \mn@doi [\apj] {10.1086/308744},
  \href {https://ui.adsabs.harvard.edu/abs/2000ApJ...534...34W} {534, 34}

\bibitem[\protect\citeauthoryear{{Wright}, {Hildebrandt}, {van den Busch}  \&
  {Heymans}}{{Wright} et~al.}{2020}]{wright20}
{Wright} A.~H.,  {Hildebrandt} H.,  {van den Busch} J.~L.,   {Heymans} C.,
  2020, \mn@doi [\aap] {10.1051/0004-6361/201936782}, \href
  {https://ui.adsabs.harvard.edu/abs/2020A&A...637A.100W} {637, A100}

\makeatother
\end{thebibliography}
\appendix

\onecolumn
\section{Demonstration of API}
\label{sec:api_appendix}

This appendix shows and explained the code needed to generate Fig.~\ref{fig:mock_data}, from the generation of the mock data corresponding to a $10^{15}$\;M$_\odot$ cluster, to the mass fitting procedure using \code{scipy} tools. When running the notebook, the user might notice warnings that some source galaxy redshifts are smaller than the cluster redshift. The code will set their lensing effects to be zero accordingly. 



Here we demonstrate how to use \code{clmm} to estimate a WL halo mass from observations of a galaxy cluster when source galaxies follow a given distribution (The LSST DESC Science Requirements Document - arXiv:1809.01669,  implemented in \code{clmm}). It uses several functionalities of the support \code{mock\_data} module to produce mock datasets.

\subsection{Generating mock data}

\code{clmm} has a support code to generate a mock catalog given a input cosmology and cluster parameters. We will use this to generate a data sample to be used in this example:

\begin{minted}[mathescape, linenos, numbersep=5pt, frame=lines, framesep=2mm]{Python}
import numpy as np
from clmm import Cosmology
import clmm.support.mock_data as mock

# For reproducibility
np.random.seed(14)

# Set cosmology of mock data
cosmo = Cosmology(H0=70.0, Omega_dm0=0.27-0.045, Omega_b0=0.045, Omega_k0=0.0)

# Cluster info
cluster_m = 1.e15 # Cluster mass - ($M200_m$) [Msun]
concentration = 4  # Cluster concentration
cluster_z = 0.3 # Cluster redshift
cluster_ra = 0. # Cluster Ra in deg
cluster_dec = 0. # Cluster Dec in deg

# Catalog info
field_size = 10 # i.e. 10 x 10 Mpc field at the cluster redshift, cluster in the center

# Make mock galaxies
mock_galaxies = mock.generate_galaxy_catalog(
    cluster_m=cluster_m, cluster_z=cluster_z, cluster_c=concentration, # Cluster data
    cosmo=cosmo, # Cosmology object
    zsrc='desc_srd', # Galaxy redshift distribution,
    zsrc_min=0.4, # Minimum redshift of the galaxies
    shapenoise=0.05, # Gaussian shape noise to the galaxy shapes
    photoz_sigma_unscaled=0.05, # Photo-z errors to source redshifts
    field_size=field_size,
    ngal_density=20 # number of gal/arcmin2 for z in [0, infty]
)['ra', 'dec', 'e1', 'e2', 'z', 'ztrue', 'pzbins', 'pzpdf', 'id']

ngals_init = len(mock_galaxies)

# Keeping only galaxies with "measured" redshift greater than cluster redshift
mock_galaxies = mock_galaxies[(mock_galaxies['z']>cluster_z)]
ngals_good = len(mock_galaxies)

if ngals_good < ngals_init:
    print(f'Number of excluded galaxies (with photoz < cluster_z): {ngals_init-ngals_good:,}')
    # reset galaxy id for later use
    mock_galaxies['id'] = np.arange(ngals_good)
\end{minted}

We can extract the column of this mock catalog to show explicitly how the quantities can be used on \code{clmm} functionality and how to add them to a \code{GalaxyCluster} object:

\begin{minted}[mathescape, linenos, numbersep=5pt, frame=lines, framesep=2mm]{Python}
# Put galaxy values on arrays
gal_ra = mock_galaxies['ra'] # Galaxies Ra in deg
gal_dec = mock_galaxies['dec'] # Galaxies Dec in deg
gal_e1 = mock_galaxies['e1'] # Galaxies elipticipy 1
gal_e2 = mock_galaxies['e2'] # Galaxies elipticipy 2
gal_z = mock_galaxies['z'] # Galaxies observed redshift
gal_ztrue = mock_galaxies['ztrue'] # Galaxies true redshift
gal_pzbins = mock_galaxies['pzbins'] # Galaxies P(z) bins
gal_pzpdf = mock_galaxies['pzpdf'] # Galaxies P(z)
gal_id = mock_galaxies['id'] # Galaxies ID
\end{minted}

\subsection{Measuring shear profiles}

From the source galaxy quantities, we can compute the ellipticities and corresponding radial profile using \code{clmm.dataops} functions:

\begin{minted}[mathescape, linenos, numbersep=5pt, frame=lines, framesep=2mm]{Python}
import clmm.dataops as da

# Convert ellipticities into shears
gal_ang_dist, gal_gt, gal_gx = da.compute_tangential_and_cross_components(cluster_ra, cluster_dec,
                                                                          gal_ra, gal_dec,
                                                                          gal_e1, gal_e2,
                                                                          geometry="flat")

# Measure profile
profile = da.make_radial_profile([gal_gt, gal_gx, gal_z],
                                 gal_ang_dist, "radians", "Mpc",
                                 bins=da.make_bins(0.01, 3.7, 50),
                                 cosmo=cosmo,
                                 z_lens=cluster_z,
                                 include_empty_bins=False)
\end{minted}

The other possibility is to use the \code{GalaxyCluster} object. This is the main approach to handle data with \code{clmm}, and also the simpler way. For that you just have to provide the following information of the cluster:

\begin{itemize}
\item Ra, Dec [deg]
\item Mass - ($M200_m$) [Msun]
\item Concentration
\item Redshift
\end{itemize}

and the source galaxies:

\begin{itemize}
\item Ra, Dec [deg]
\item 2 axis of eliptticities
\item Redshift
\end{itemize}

\begin{minted}[mathescape, linenos, numbersep=5pt, frame=lines, framesep=2mm]{Python}
import clmm

# Create a GCData with the galaxies
galaxies = clmm.GCData([gal_ra, gal_dec, gal_e1, gal_e2, gal_z,
                        gal_ztrue, gal_id],
                      names=['ra', 'dec', 'e1', 'e2', 'z',
                             'ztrue', 'id'])

# Create a GalaxyCluster
cluster = clmm.GalaxyCluster("Name of cluster", cluster_ra, cluster_dec,
                                   cluster_z, galaxies)

# Convert elipticities into shears for the members
cluster.compute_tangential_and_cross_components(geometry="flat")

# Measure profile and add profile table to the cluster
cluster.make_radial_profile(bins=da.make_bins(0.1, field_size/2., 25, method='evenlog10width'),
                            bin_units="Mpc",
                            cosmo=cosmo,
                            include_empty_bins=False,
                            gal_ids_in_bins=True,
                           )
\end{minted}

This results in an attribute \code{table} added to the \code{cluster} object.

\subsection{Theoretical predictions}

We consider 3 models:
1. One model where all sources are considered at the same redshift
2. One model using the overall source redshift distribution to predict the reduced tangential shear
3. A more accurate model, relying on the fact that we have access to the individual redshifts of the sources, where the average reduced tangential shear is averaged independently in each bin, accounting for the acutal population of sources in each bin.

All models rely on \code{clmm.predict\_reduced\_tangential\_shear} to make a prediction that accounts for the redshift distribution of the galaxies in each radial bin:

\subsubsection{Model considering all sources located at the average redshift}
\begin{equation*}
     g_{t,i}^{\rm{avg(z)}} = g_t(R_i, \langle z \rangle)\;,
 \label{eq:wrong_gt_model}
 \end{equation*}

\begin{minted}[mathescape, linenos, numbersep=5pt, frame=lines, framesep=2mm]{Python}
def predict_reduced_tangential_shear_mean_z(profile, logm):
    return clmm.compute_reduced_tangential_shear(
            r_proj=profile['radius'], # Radial component of the profile
            mdelta=10**logm, # Mass of the cluster [M_sun]
            cdelta=4, # Concentration of the cluster
            z_cluster=cluster_z, # Redshift of the cluster
            z_source=np.mean(cluster.galcat['z']), # Mean value of source galaxies redshift
            cosmo=cosmo,
            delta_mdef=200,
            halo_profile_model='nfw'
        )
\end{minted}

\subsubsection{Model relying on the overall redshif distribution of the sources, not using individual redshift information, i.e. approximate  (from \citealt{applegate14})}
\begin{equation*}
     g_{t,i}^{N(z)} = \frac{\langle\beta_s\rangle \gamma_t(R_i, z\rightarrow\infty)}{1-\frac{\langle\beta_s^2\rangle}{\langle\beta_s\rangle}\kappa(R_i, z\rightarrow\infty)}
     \label{eq:approx_model}
 \end{equation*}

\begin{minted}[mathescape, linenos, numbersep=5pt, frame=lines, framesep=2mm]{Python}
z_inf = 1000
dl_inf = cosmo.eval_da_z1z2(cluster_z, z_inf)
d_inf = cosmo.eval_da(z_inf)

def betas(z):
    dls = cosmo.eval_da_z1z2(cluster_z, z)
    ds = cosmo.eval_da(z)
    return dls * d_inf / (ds * dl_inf)

def predict_reduced_tangential_shear_approx(profile, logm):

    bs_mean = np.mean(betas(cluster.galcat['z']))
    bs2_mean = np.mean(betas(cluster.galcat['z'])**2)

    gamma_t_inf = clmm.compute_tangential_shear(
            r_proj=profile['radius'], # Radial component of the profile
            mdelta=10**logm, # Mass of the cluster [M_sun]
            cdelta=4, # Concentration of the cluster
            z_cluster=cluster_z, # Redshift of the cluster
            z_source=z_inf, # Redshift value at infinity
            cosmo=cosmo,
            delta_mdef=200,
            halo_profile_model='nfw')
    convergence_inf = clmm.compute_convergence(
            r_proj=profile['radius'], # Radial component of the profile
            mdelta=10**logm, # Mass of the cluster [M_sun]
            cdelta=4, # Concentration of the cluster
            z_cluster=cluster_z, # Redshift of the cluster
            z_source=z_inf, # Redshift value at infinity
            cosmo=cosmo,
            delta_mdef=200,
            halo_profile_model='nfw')

    return bs_mean*gamma_t_inf/(1-(bs2_mean/bs_mean)*convergence_inf)
\end{minted}

\subsubsection{Model using individual redshift and radial information, to compute the averaged shear in each radial bin, based on the galaxies actually present in that bin.}
\begin{equation*}
    g_{t,i}^{z, R} = \frac{1}{N_i}\sum_{{\rm gal\,}j\in {\rm bin\,}i} g_t(R_j, z_j)
    \label{eq:exact_model}
 \end{equation*}

\begin{minted}[mathescape, linenos, numbersep=5pt, frame=lines, framesep=2mm]{Python}
cluster.galcat['theta_mpc'] = convert_units(
    cluster.galcat['theta'], 'radians', 'mpc',
    cluster.z, cosmo)
def predict_reduced_tangential_shear_exact(profile, logm):

    return np.array([np.mean(
        clmm.compute_reduced_tangential_shear(
            # Radial component of each source galaxy inside the radial bin
            r_proj=cluster.galcat[radial_bin['gal_id']]['theta_mpc'],
            mdelta=10**logm, # Mass of the cluster [M_sun]
            cdelta=4, # Concentration of the cluster
            z_cluster=cluster_z, # Redshift of the cluster
            # Redshift value of each source galaxy inside the radial bin
            z_source=cluster.galcat[radial_bin['gal_id']]['z'],
            cosmo=cosmo,
            delta_mdef=200,
            halo_profile_model='nfw'
        )) for radial_bin in profile])
\end{minted}

\subsection{Mass fitting}

We estimate the best-fit mass using \code{scipy.optimize.curve\_fit}. The choice of fitting $\log M$ instead of $M$ lowers the range of pre-defined fitting bounds from several order of magnitude for the mass to unity. From the associated error $\sigma_{\log M}$ we calculate the error to mass as $\sigma_M = M_{fit}\ln(10)\sigma_{\log M}$.

First, we identify bins with sufficient galaxy statistics to be kept for the fit.
For small samples, error bars should not be computed using the simple error on the mean approach available so far in CLMM.

\begin{minted}[mathescape, linenos, numbersep=5pt, frame=lines, framesep=2mm]{Python}
data_for_fit = cluster.profile[cluster.profile['n_src']>5]
\end{minted}

Fitting the data:

\begin{minted}[mathescape, linenos, numbersep=5pt, frame=lines, framesep=2mm]{Python}
from clmm.support.sampler import fitters
# Single z value model
popt, pcov = fitters['curve_fit'](predict_reduced_tangential_shear_mean_z,
    data_for_fit, data_for_fit['gt'],
    data_for_fit['gt_err'], bounds=[10.,17.])
logm_1z, logm_1z_err = popt[0], np.sqrt(pcov[0][0])
# Applegate et at. 2014
popt, pcov = fitters['curve_fit'](predict_reduced_tangential_shear_approx,
    data_for_fit, data_for_fit['gt'],
    data_for_fit['gt_err'], bounds=[10.,17.])
logm_ap, logm_ap_err = popt[0], np.sqrt(pcov[0][0])
# Considering all redshifts and radii
popt, pcov = fitters['curve_fit'](predict_reduced_tangential_shear_exact,
    data_for_fit, data_for_fit['gt'],
    data_for_fit['gt_err'], bounds=[10.,17.])
logm_exact, logm_exact_err = popt[0], np.sqrt(pcov[0][0])
\end{minted}

\end{document}